\documentclass[letterpaper,12pt,draftcls,onecolumn]{IEEEtran}

\usepackage{amsmath}
\usepackage{amssymb}
\usepackage{amsthm}
\usepackage{graphicx}
\usepackage{accents}
\usepackage[noadjust]{cite}
\usepackage{color}

\newtheorem{theorem}{Theorem}
\newtheorem{lemma}[theorem]{Lemma}
\newtheorem{corollary}[theorem]{Corollary}

\newcommand{\ubar}[1]{\underaccent{\bar}{#1}}

\newcommand{\Lie}[1]{\mathcal{L}_{#1}}
\newcommand{\norm}[1]{\left\lVert#1\right\rVert}

\newcommand{\eps}{\varepsilon}
\newcommand{\real}{\mathbb{R}}

\DeclareMathOperator{\rank}{rank}
\DeclareMathOperator{\tr}{tr}
\DeclareMathOperator{\E}{E}
\DeclareMathOperator{\Var}{Var}
\DeclareMathOperator{\Cov}{Cov}
\DeclareMathOperator{\vecop}{vec}
\DeclareMathOperator{\diag}{diag}

\title{Empirical Observability Gramian for Stochastic Observability of Nonlinear Systems}
\author{Nathan~D.~Powel,~\IEEEmembership{Member,~IEEE} and~Kristi~A.~Morgansen,~\IEEEmembership{Senior~Member,~IEEE}%
\thanks{This work was funded by in part by ONR MURI N000141010952.}
\thanks{The authors are with the Department of Aeronautics \& Astronautics, University of Washington, Seattle, WA 98195-2400. \{{\tt\small poweln,morgansen}\}{\tt\small @aa.washington.edu}.}
}

\begin{document}

\markboth{IEEE Transactions on Automatic Controls,~Vol.~X, No.~Y, Month~Year}%
{Powel \MakeLowercase{\textit{et al.}}: Empirical Observability Gramian for Stochastic Observability of Nonlinear Systems}

\maketitle
\thispagestyle{empty}
\pagestyle{empty}

\begin{abstract}
We extend observability metrics based on the empirical observability Gramian from deterministic nonlinear systems to nonlinear stochastic systems in order to capture the impact of process noise on observability.  We demonstrate that the empirical observability Gramian can be used to provide an equivalent condition for a definition of stochastic observability on linear systems, and that the Gramian can be used to extend stochastic observability to nonlinear stochastic systems.  We further demonstrate through simulation that consideration of process noise can reveal observability in systems that would be considered unobservable using traditional deterministic tools.
\end{abstract}


\section{Introduction}
\label{sec:introduction}

\IEEEPARstart{W}{hile} the importance of control in the analysis of observability of nonlinear systems is well understood, the influence of process noise on observability is less well studied.  Process noise is known excite meaningful non-stochastic effects in nonlinear systems such as stochastic resonance in climate models \cite{Benzi1982,Nicolis1982} to enhanced signal transmission in neural systems \cite{Godivier1996}.  Stochastic inputs are widely used in system identification and adaptive control to provide persistently exciting actuation.  Because noise is ubiquitous in physically instantiated systems (arising from phenomena such as aerodynamic turbulence, electrical noise, thermal fluctuations and other unmodeled or incompletely understood physics \cite{Longtin2003}), we believe that a better understanding of the influence of process noise on observability is needed to determine if the beneficial effects of noise can, in some circumstances, include enhanced observability as well.  In this work we propose to extend the empirical observability Gramian to stochastic nonlinear systems and to use the Gramian as a tool to investigate the effect of process noise on the observability of nonlinear systems. 

To illustrate how actuation by process noise might improve observability, imagine a stationary planar unicycle vehicle on which we can observe only the position.  If we wish to observe the heading of this vehicle, we must apply an acceleration input.  There is no reason, however, that such an input must be deterministic.  Process noise in the acceleration/speed dynamics could serve, as well, to determine the orientation of the vehicle ($\pm180^\circ$ due to symmetry in the dynamics; heading would be only locally observable).  Because noise is capable of actuating states that control input cannot, in some systems, noise must be considered when determining observability.

Many competing notions of observability exist for stochastic systems, though many have been defined and examined only in the case of linear stochastic systems \cite{Aoki1968,Sunahara1975,Chen1980,Mariton1986,Baram1987,Gustafsson2001,Hwang2003,Dragan2004,Zhang2004,Sun2010,Ting2012,Subasi2014}, and the community does not seem to have achieved consensus on a preferred approach.  Many of these definitions of stochastic observability follow common themes, including definitions based on attempts to extend indistinguishability to probability distributions of the state \cite{Hwang2003}, definitions based on convergence of state estimates or their covariance below a particular threshold \cite{Aoki1968,Sunahara1974,Sunahara1975a,Sunahara1975,Bageshwar2009}, stochastic controllability of the dual linear system \cite{Mariton1986}, mutual information between the state and the output \cite{Baram1987,West1994,Liu2009a,Liu2010,Liu2011,Liu2011a,Shen2011,Subasi2014,Zhang2015b}, and generalizations of deterministic exact observability/detectability \cite{Li2009,Zhang2004,Zhang2008a,Sun2010,Ting2012,Shen2013} (note that some of these definitions are joint properties of a system and a chosen estimator).  Related work has also examined the observability of uncertain linear systems \cite{Ugrinovskii2002} with noise.

For nonlinear systems, Liu and Bitmead \cite{Liu2009a,Liu2010,Liu2011,Liu2011a} examine estimability, (and a specialization, \emph{complete reconstructability}, which reduces to complete reconstructability of deterministic linear systems) for linear and nonlinear discrete-time stochastic systems.  However, the estimability of nonlinear systems requires that every measurable function of the initial state with non-zero entropy must have strictly positive mutual information with the measurement, a property which is extremely difficult to check for most nonlinear systems.  Beyond the work of Liu and Bitmead, comparatively little effort has been expended examining the observability of nonlinear stochastic systems.  In the mid-1970's, Sunahara et al. defined stochastic observability for nonlinear systems (in continuous and discrete-time) based on a probability threshold of a linear feedback estimator converging to within specified error \cite{Sunahara1974,Sunahara1975a,Sunahara1975}.  The metric of stochastic observability used by Subasi and Demirekler \cite{Subasi2014} applies to nonlinear systems, but all analysis using that metric was restricted to LTI systems.  And while some of the definitions of stochastic observability could potentially be extended to nonlinear stochastic systems, definitions that depend explicitly on linear system matrices \cite{Gustafsson2001,Dragan2004,Dragan2006} or that assume a particular form of estimator \cite{Sunahara1974,Sunahara1975a,Sunahara1975,Bageshwar2009} do not extend as naturally to general nonlinear systems.  Nonlinear systems are more interesting from an observability perspective because of the lack of a separation principle.

Here, we approach observability in nonlinear stochastic systems using the empirical observability Gramian.  The empirical observability Gramian, unlike existing stochastic observability concepts, provides a way forward with a strong connection to observability in deterministic nonlinear systems, and which is amenable to \emph{tractable} application to arbitrary nonlinear stochastic systems \cite{Powel2015}.  The empirical observability Gramian was originally introduced by Lall, Marsden, and Glava\v{s}ki \cite{Lall1999} and initially used in state-space reduction of nonlinear systems \cite{Lall1999,Lall2002,Hahn2002}.  The empirical observability Gramian is an extension of the linear observability Gramian to nonlinear systems, with the advantage that only the ability to simulate the system is needed to compute it; the nonlinear system need not be analytically tractable, or even known in closed-form.

The empirical observability Gramian has since been applied as a tool for acquiring quantitative metrics of the observability of nonlinear systems \cite{Singh2005,Krener2009}.  The local unobservability index, and the estimation condition number, derived from the Gramian, identified by Krener and Ide \cite{Krener2009} as useful nonlinear observability metrics, have seen particularly wide adoption.  Application areas have included underwater navigation \cite{Hinson2013,Glotzbach2014}, planetary landers \cite{Yu2015}, blood glucose modeling \cite{Eberle2012}, and flow-field estimation \cite{Hinson2012,DeVries2013}.  The metrics have also been adapted to analyzing the observability of subspaces of systems of partial differential equations \cite{Kang2014,King2014}.

Until recently, use of the empirical observability Gramian and the observability metrics associated with it have been largely heuristically justified.  Rigorous connections to observability have emerged only recently.  Batista, Silvestre, and Oliveira \cite{Batista2011} demonstrated that the rank of observability Gramian could be used to prove observability of a class of approximately linear time-varying nonlinear systems.  In previous work, Powel and Morgansen \cite{Powel2015} developed a sufficient condition for the weak observability of an arbitrary, deterministic, nonlinear system, with control, based upon the empirical observability Gramian.

In this paper, we build upon results from \cite{Powel2015} to extend the empirical observability Gramian as a tool to analyze observability of stochastic nonlinear systems.  In particular, we demonstrate that the empirical observability Gramian can provide a unified approach to the observability analysis of deterministic and stochastic nonlinear systems by providing a test for the Dragan and Morozan definition of stochastic observability for linear stochastic systems with multiplicative noise \cite{Dragan2004}.  The empirical observability Gramian can, therefore, be used to extend that definition to nonlinear stochastic systems as well.  We also demonstrate the numerical computation of our previously derived lower bound on the minimum singular value of the Gramian and demonstrate the impact of process noise on the observability metrics proposed by \cite{Krener2009}.  Although many of our results for deterministic systems can be trivially adapted to discrete time systems by switching from integration to summation, we operate on continuous time systems in this paper to provide continuity moving from determistic systems to stochastic processes.

The remainder of the paper is organized as follows.  Section~\ref{sec:background} provides some important mathematical preliminaries to the results of this paper.  Section~\ref{sec:gramianobsv} contains theorems connecting the minimum singular value of empirical observability Gramian to weak observability of deterministic nonlinear systems.  In Section~\ref{sec:noisegram} we introduce an extension of the empirical observability Gramian to stochastic processes and derive some properties of the extended Gramian.  We also prove a rank condition on the first moment of the Gramian of a stochastic linear system with multiplicative noise that is equivalent to stochastic observability.  Section~\ref{sec:err} applies the noise extended Gramian to the case of noise arising from modeling error, and in Section~\ref{sec:numeric} we numerically evaluate the Gramian for some sample stochastic systems.  We wrap up the paper in Section~\ref{sec:conclusions} with conclusions and future work.

\section{Preliminaries}
\label{sec:background}

In this paper we will discuss two different kinds of dynamical system.  The first kind, nonlinear deterministic systems, we write as the ordinary differential equation (ODE)
\begin{equation}
\label{eq:system}
\Sigma_D:\quad
\begin{aligned}
\dot x & = f(x,u)\\
y & = h(x),
\end{aligned}
\end{equation}
where $x(t) \in \real^n$, $u(t) \in \real^m$, $y(t) \in \real^p$, $u \in \mathcal{U}$ where $\mathcal{U}$ is the set of permissible controls.  We will write the solution to the initial value problem for \eqref{eq:system} for $x(0) = x_0$ with the control input $u(t)$ as $x(t, x_0, u)$ and write $y(t,x_0,u) = h(x(t,x_0,u))$.  We will also assume that $f,h \in C^\infty$, which is sufficient to guarantee existence and uniqueness of a solution to \eqref{eq:system} (at least for well-behaved $u(t)$ and over some non-empty time interval), as well as to allow us to apply Taylor's theorem.

The second kind of system, nonlinear stochastic systems, we will write as a stochastic differential equation (SDE)
\begin{equation}
\label{eq:sde}
\Sigma_S:\quad
\begin{aligned}
dX & = f(X,u) dt + \sigma(X,u) dW\\
Y & = h(X),
\end{aligned}
\end{equation}
where $dW$ is a vector of independent differentials of the It\^{o} sense.  The same assumptions as to the continuity of $f(X,u)$, $\sigma(X,u)$, and $h(X)$ will be made as for the components of the deterministic system.  However, $X$ now represents a vector of random variables over $\real^n$, and, in general, there is no longer a unique state trajectory in the set of mappings from $\real$ to $\real^n$ that satisfies \eqref{eq:sde}.  The probability distribution of $X(t)$ has deterministic dynamics given by the Fokker-Planck equations and sample trajectories of the system can be drawn satisfying the probability distribution by methods such as the Euler-Maruyama method or the Milstein Method.

Noise in the output of the system cannot influence the state trajectory, meaning that measurement noise is incapable of increasing the observability of a system.  To simplify the analysis, we will therefore neglect measurement noise in this paper.

\subsection{Weak observability}

In order to be certain that our terminology is clear, we will briefly outline how observability is defined for nonlinear deterministic systems, and give sufficient conditions for deterministic nonlinear systems $\Sigma_D$ to meet these definitions.  The definitions used in this section originate in \cite{Hermann1977}.  Further details on the Lie algebraic approach to observability of nonlinear deterministic systems can be found in \cite{Hermann1977} and in \cite{Nijmeijer1990}.

We say that points $x_0, x_1 \in \real^n$ are \emph{indistinguishable} if for every control, $u \in \mathcal{U}$, $y(t,x_0,u) = y(t,x_1,u)$ for all $t$.  We say that $\Sigma_D$ is \emph{weakly observable at $x_0$} if there exists an open neighborhood $U$ of $x_0$ such that if $x_1 \in U$ and $x_0$ and $x_1$ are indistinguishable, then $x_0 = x_1$, and we say that $\Sigma_D$ is \emph{weakly observable} if $\Sigma_D$ is weakly observable at all $x$.

We say that points $x_0$ and $x_1$ are \emph{$U$-indistinguishable} if for every control, $u \in \mathcal{U}$, with trajectories $x(t,x_0,u)$ and $x(t,x_1,u)$ that lie in $U \subseteq \real^n$ for $t \in [0, T]$, we have $y(t,x_0,u) = y(t,x_1,u)$ for all $t \in [0,T]$.  We say that $\Sigma_D$ is \emph{locally weakly observable at $x_0$} if there exists an open neighborhood $U$ of $x_0$ such that for every open neighborhood $V \subset U$ of $x_0$, $x_0$ and $x_1$ $V$-indistinguishable implies that $x_0 = x_1$, and we say that $\Sigma_D$ is \emph{locally weakly observable} if $\Sigma_D$ is locally weakly observable at all $x$.

Note that local weak observability implies weak observability.  Intuitively, weak observability at $x_0$ implies that $x_0$ can be eventually distinguished from its neighbors for some control, while local weak observability implies that $x_0$ can be instantly distinguished from its neighbors for some control \cite{Hermann1977}.

The usual approach to testing the observability of a nonlinear system comes from differential geometry, and provides a rank condition for local weak observability.  We define the \emph{Lie derivative} of the function, $h(x)$, with respect to a vector field, $f_i(x)$, as
\begin{equation}
\Lie{f_i(x)} h(x) = f_i(x)^T \frac{\partial h}{\partial x}.
\end{equation}
Because the result of a Lie derivative operation is a vector field mapping between the same two spaces, Lie derivatives can be applied sequentially.

The \emph{observability Lie algebra}, $\mathcal{O}$, of a system $\Sigma$, is the span of the Lie derivatives of the output function, $h$:
\begin{equation}
\mathcal{O}(x) = \operatorname{span} \{ \Lie{X_1}\Lie{X_2} \ldots \Lie{X_k} h(x) \},
\end{equation}
where $X_i \in \{f(x,u_0) \mid u_0 \in \mathcal{U}\}$ for $i \in \{1, 2, \ldots, k\}$.  If the Jacobian of any set of vectors in the observability Lie algebra, $d\mathcal{O}(x_0)$, is full rank at some state $x_0$, then the system is locally weakly observable at $x_0$ \cite{Hermann1977}.  If the system is control affine (i.e. $\dot x = f_0(x) + \sum_{i = 1}^m f_i(x)u_i$) then $X_i \in \{f_0, f_1, \ldots, f_m\}$ for $i \in \{1, 2, \ldots, k\}$ \cite{Nijmeijer1990}.

Note that these definitions of observability do not map well to stochastic systems, because even for a deterministic initial condition (initial condition having a Dirac distribution) a stochastic system can have multiple sample measurement trajectories.  In other words, strictly applying the definitions above to stochastic systems, a given point is not guaranteed to be indistinguishable from itself.  Unlike both linear and nonlinear deterministic systems, there is no single universally accepted definition for stochastic observability that maps well to weak observability.  However, as we will show later in Section~\ref{sec:noisegram}, the empirical observability Gramian itself can be used to define a notion of observability for stochastic nonlinear systems.

Of course process noise is also substantially different from control input.  Noise is infinitely discontinuous, and does not have a known magnitude, or even sign (though the statistics of both of these are generally assumed to be known).  This property makes the noise much more difficult to use in conjunction with an estimator, and can make it difficult to distinguish between certain kinds of symmetrical dynamics.

\subsection{Empirical observability Gramian}

We define the empirical observability Gramian for $\Sigma_D$ as
\begin{equation}
\label{eq:gramian}
W_o^\eps(t_1, x_0, u) = \frac{1}{4\eps^2} \int_0^{t_1} \Phi^\eps(t,x_0,u)^T \Phi^\eps(t,x_0,u) dt,
\end{equation}
where
\begin{equation}
\label{eq:phi}
\Phi^\eps(t,x_0,u) = \begin{bmatrix}y^{+1}-y^{-1} & \dotsc & y^{+n}-y^{-n}\end{bmatrix}
\end{equation}
and
\begin{equation}
\label{eq:ypm}
y^{\pm i} = y(t, x_0 \pm \eps e_i, u).
\end{equation}
The vectors $e_i$ denote the elements of the standard basis in $\real^n$.  Note that this definition differs slightly from that in \cite{Krener2009} by explicitly including control.  The empirical observability Gramian reduces to the well-known linear observability Gramian when $\Sigma_D$ is a linear deterministic system.  Krener and Ide \cite{Krener2009} introduce two numbers based on the empirical observability Gramian to provide quantitative measures of nonlinear observability.  The local unobservability index is the reciprocal of the minimum eigenvalue of the Gramian and the estimation condition number is the condition number of the Gramian.  For both numbers, lower values indicate greater observability.

In the case of linear deterministic systems, a system is observable if and only if the Gramian has full rank.  In \cite{Powel2015} we showed that an arbitrary nonlinear system \eqref{eq:system} is weakly observable when the unobservability index is below a certain upper bound.  In this work, we will show that the full rank of the first moment of the empirical observability Gramian is equivalent to the stochastic observability of a linear stochastic system with multiplicative noise.


\section{Deterministic nonlinear observability}
\label{sec:gramianobsv}

In previous work, \cite{Powel2015}, we showed that the empirical observability Gramian could be used to determine weak observability for nonlinear systems.  We will now review the major results of that work prior to discussing how they can be numerically computed, and to provide a context for the stochastic extensions that we make in Section~\ref{sec:noisegram}.

We prove that in the limit as $\eps \to 0$, the system is weakly observable if the Gramian is full rank
\begin{theorem}
\label{thm:weakobsv}
If there exists $u \in \mathcal{U}$ such that
\begin{equation}
\rank\left(\lim_{\eps \to 0} W_o^\eps(t_1, x_0, u)\right) = n
\end{equation}
for some $t_1 > 0$, then the system is weakly observable at $x_0$.
\end{theorem}
\begin{IEEEproof}
See \cite{Powel2015}.
\end{IEEEproof}

While this result is interesting, the Gramian above is rarely analytically computable.  However, by computing the error of the finite $\eps$ Gramian from the limit Gramian, we also proved 
\begin{theorem}
\label{thm:empweakobsv}
If there exists $u \in \mathcal{U}$ such that
\begin{equation}
\label{eq:bound}
\ubar{\sigma}(W_o^\eps) > \sup_{t \in [0, t_1]}\left( \frac{\sqrt{n} \eps^2 t_1}{3} \norm{\frac{\partial y}{\partial x_0}}_2 \Gamma + \frac{n \eps^4 t_1}{36} \Gamma^2 \right)
\end{equation}
for some $t_1 > 0$, where
\begin{equation}
\Gamma(t,x_0,u) = \max_i \sup_{\eta \in \mathcal{I}_i^\eps} \norm{D^3y(\eta)(e_i, e_i, e_i)}_1,
\end{equation}
and $\mathcal{I}_i^\eps = [x_0 - \eps e_i, x_0 + \eps e_i]$ is the closed line segment from $x_0 - \eps e_i$ to $x_0 + \eps e_i$, then the system is weakly observable at $x_0$.
\end{theorem}
\begin{IEEEproof}
See \cite{Powel2015}.
\end{IEEEproof}

In general, the lower bound in Theorem~\ref{thm:empweakobsv} is rarely analytically computable, because $y(t)$ is not generally an analytical function of $x_0$ for $t > 0$ and, as a result, we cannot compute $\frac{\partial y}{\partial x_0}$.  However, we can approximately compute the lower bound by using a finite differencing method to compute $\Gamma$ and $\frac{\partial y}{\partial x_0}$.  Any such lower bound is approximate, and cannot be used to rigorously prove a system weakly observable, but it does provide a reasonable heuristic for evaluating the index of unobservability that we numerically compute for stochastic systems below.

In this work, we approximate $\frac{\partial y}{\partial x_0}$ and $D^3y(\eta)(e_i, e_i, e_i)$ with second order central difference methods.  The partial derivative $\frac{\partial y}{\partial x_0}$ we get from
\begin{equation}
\frac{\partial y}{\partial x_0}(t_k) \approx \frac{\Phi^\eps(t_k)}{2 \eps}.
\end{equation}
To get $D^3y(\eta)(e_i, e_i, e_i)$ we discretize the interval $\mathcal{I}_i^\eps$ into $N$ points $x_{i,j}$ for $j \in \{1, \dotsc, N\}$ and compute $D^3y(\eta)(e_i, e_i, e_i)$ at each point.  Define
\begin{equation}
y^{++i}(t, x_{i,j}, u) = h(x(t, x_{i,j} + 2 dx e_i, u))
\end{equation}
and
\begin{equation}
y^{--i}(t, x_{i,j}, u) = h(x(t, x_{i,j} - 2 dx e_i, u))
\end{equation}
for some small $dx$.  Then we can compute the Fr\'{e}chet derivative of $y$ by simulating the system trajectory four times for each point and calculating
\begin{equation}
\begin{split}
D^3y(x_{i,j})(e_i, e_i, e_i) \approx & \frac{1}{dx^3}\left(\frac{1}{2} y^{++i}(t_k,x_{i,j}) - y^{+i}(t_k,x_{i,j})\right.\\
&\left. + y^{-i}(t_k,x_{i,j}) - \frac{1}{2} y^{--i}(t_k,x_{i,j})\right).
\end{split}
\end{equation}
From here it is a simple, but computationally intensive, matter to compute an approximate lower bound from \eqref{eq:bound}.

Performing this procedure for an observable linear system
\begin{equation}
\begin{aligned}
\dot x & = \begin{bmatrix}-x_2 \\ x_1\end{bmatrix}\\
y & = x_2.
\end{aligned}
\end{equation}
we find that the minimum singular value of $W_o^\eps(10,0,0)$ must exceed $1.8 \times 10^{-4}$ to guarantee weak observability.  We expect that for a linear system, the empirical observability Gramian should approximately equal the actual observability Gramian, so for an observable linear system the bound that we expect to get should be approximately zero, which is what we found.  In this case, the empirical observability Gramian actually has a minimum singular value of approximately $4.7$, indicating that the system is indeed observable.  If we change the linear system to make it unobservable, 
\begin{equation}
\begin{aligned}
\dot x & = \begin{bmatrix}-x_2 \\ 0\end{bmatrix}\\
y & = x_2.
\end{aligned}
\end{equation}
we find, that the minimum singular value of the empirical observability Gramian is now $0$.

We can also perform this analysis for a nonlinear unicycle system with the dynamics
\begin{equation}
\begin{aligned}
\dot x & =
\begin{bmatrix}
x_4 \cos(x_3)\\
x_4 \sin(x_3)\\
u_1\\
u_2
\end{bmatrix}
\\
y & =
\begin{bmatrix}
x_1\\
x_2
\end{bmatrix}.
\end{aligned}
\end{equation}
With the system at the origin, the numerical lower bound for observability on the Gramian singular value is zero, and with no control input, the Gramian has a singular value of zero.  Adding a control acceleration of $u(t) = 1$, we get a singular value of $2.44$, which exceeds the required singular bound of $1.68$ (recall that the bound itself depends on the integration time, initial condition, and control input), indicating that the system is observable at the origin only with control.

In \cite{Powel2015}, a connection between the Fisher information matrix and the empirical observability Gramian was demonstrated.

\begin{theorem}
\label{thm:fisher}
For a nonlinear system with
\begin{equation}
\label{eq:fisher}
\tilde y = h(x) + v,
\end{equation}
where $v \sim \mathcal{N}(0,R)$, we can bound the Fisher information of $y(t, x_0, u)$ with respect to $x_0$ by
\begin{equation}
F(t) \preccurlyeq \bar\sigma\left(R^{-1}\right) \frac{d}{dt} \lim_{\eps \to 0} W_o^\eps(t, x_0, u),
\end{equation}
where $\bar{\sigma}(R^{-1})$ and $\ubar{\sigma}(R^{-1})$ denote the maximum and minimum singular values of $R^{-1}$ respectively, and $\preccurlyeq$ refers to the positive semidefinite order relation for square matrices.
\end{theorem}
\begin{IEEEproof}
See \cite{Powel2015}.
\end{IEEEproof}

Theorem~\ref{thm:fisher} suggests that the shape of the empirical observability Gramian can be useful in determining the likely performance limits and conditioning of nonlinear estimators for our system around a particular state.  Because the Fisher information is bounded above and below by scalings of the integrand of the empirical Gramian, when the estimation condition number of the system is high (and particularly when the condition number of $R^{-1}$ is also low), the Fisher information is also likely to have a high condition number, which in turn places constraints on the numerical conditioning of unbiased estimators applied to the problem.  We can formalize this in the following corollary to Theorem~\ref{thm:fisher}.

\begin{corollary}
\label{thm:fishercond}
For the system given by \eqref{eq:fisher},
\begin{equation}
\max \left\{ 1, \frac{ \kappa\left(\frac{d}{dt} \lim_{\eps \to 0} W_o^\eps \right)}{\kappa(R)} \right\} \leq \kappa(F) \leq \kappa(R) \kappa\left(\frac{d}{dt} \lim_{\eps \to 0} W_o^\eps\right)
\end{equation}
%
%
where $\kappa(A)$ is the condition number of the matrix $A$.
\end{corollary}
\begin{IEEEproof}
From Theorem~\ref{thm:fisher}, we know that
\begin{equation}
\bar \lambda(F) \leq \bar \lambda(R^{-1}) \bar \lambda\left(\frac{d}{dt} \lim_{\eps \to 0} W_o^\eps\right)
\end{equation}
and
\begin{equation}
\ubar \lambda(F) \geq \ubar \lambda(R^{-1}) \ubar \lambda\left(\frac{d}{dt} \lim_{\eps \to 0} W_o^\eps\right),
\end{equation}
using the fact the $\lambda_i(R^{-1}) = \sigma_i(R^{-1})$ because $R \succ 0$.  It follows immediately that
\begin{equation}
\kappa(F) = \frac{\bar \lambda(F)}{\ubar \lambda(F)} \leq \frac{\bar \lambda(R^{-1}) \bar \lambda\left(\frac{d}{dt} \lim_{\eps \to 0} W_o^\eps\right)}{\ubar \lambda(R^{-1}) \ubar \lambda\left(\frac{d}{dt} \lim_{\eps \to 0} W_o^\eps\right)} = \kappa(R) \kappa\left(\frac{d}{dt} \lim_{\eps \to 0} W_o^\eps\right),
\end{equation}
%
%
%
because the condition number of a positive-definite matrix and its inverse are the same.

To arrive at the other part of the inequality, let us assume that $F = \bar\lambda(R^{-1}) \left(\frac{d}{dt} \lim_{\eps \to 0} W_o^\eps\right)$.  This choice of $F$ fits the bounds of Theorem~\ref{thm:fisher} and has the largest possible minimum eigenvalue, $\ubar \lambda(F) = \bar \lambda(R^{-1}) \ubar \lambda\left(\frac{d}{dt} \lim_{\eps \to 0} W_o^\eps\right)$.  Now we smoothly reduce the maximum eigenvalue of $F$, reducing the condition number of $F$, until either $\bar \lambda(F) = \ubar \lambda(F)$ , in which case $\kappa(F) = 1$, its smallest possible value, or until we run into the lower bound from Theorem~\ref{thm:fisher}, $\bar \lambda(F) = \ubar \lambda(R^{-1}) \bar \lambda \left(\lim_{\eps \to 0} W_o^\eps\right)$.  Thus, we have
\begin{equation}
\max \left\{ 1, \frac{\kappa \left(\frac{d}{dt} \lim_{\eps \to 0} W_o^\eps \right)}{\kappa(R)} \right\} = \max \left\{ 1, \frac{\ubar \lambda(R) \bar \lambda\left(\frac{d}{dt} \lim_{\eps \to 0} W_o^\eps \right)}{\bar \lambda(R) \ubar \lambda\left(\frac{d}{dt} \lim_{\eps \to 0} W_o^\eps \right)} \right\} \leq \frac{\bar \lambda(F)}{\ubar \lambda(F)} = \kappa(F)
\end{equation}
%
%
We could also have begun with $F = \ubar\lambda(R^{-1}) \left(\frac{d}{dt} \lim_{\eps \to 0} W_o^\eps\right)$ and smoothly increased the minimum eigenvalue of $F$ to arrive at the same result.
\end{IEEEproof}

\begin{figure}
\centering
\includegraphics[width=0.5\textwidth]{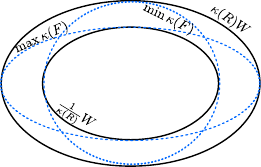}
\caption[Fisher information matrix condition number bounds]{The maximum and minimum of the condition number of the Fisher information matrix are bounded by the condition number of the integrand of the empirical observability Gramian scaled by the condition number of the covariance of the measurement noise ($W$ stands for $\frac{d}{dt} \lim_{\eps \to 0} W_o^\eps$).}
\label{fig:fisher}
\end{figure}

Figure~\ref{fig:fisher} illustrates the intuition behind Corollary~\ref{thm:fishercond}.  The Fisher information matrix ellipsoid is constrained to remain between the two scalings of the ellipsoid given by the integrand of the Gramian.  The larger the condition number of the measurement noise covariance, the further apart those ellipsoids will lie, and the larger the freedom there is in the condition number of the Fisher information.  If the condition number of $R$ is unity, then there is no room between the ellipsoids at all, and the inequalities of Theorem~\ref{thm:fisher} become equality.

Note that $\kappa\left(\lim_{\eps \to 0} W_o^\eps\right)$ is the estimation condition number of the empirical observability Gramian in the limit as $\eps \to 0$.  Thus, the conditioning of the Fisher information matrix is bounded above by a scaling of the estimation condition number.  The closer the condition number of the measurement noise covariance is to unity, the tighter the connection between the estimation condition number and the Fisher information matrix.  Thus, we can rigorously connect the numerical conditioning of the estimation problem to the other metric of observability from \cite{Krener2009}.

\section{Stochastic Gramian}
\label{sec:noisegram}

We now examine how we can incorporate process noise into the empirical observability Gramian framework.  The most straightforward approach is to add process noise to the dynamics of the system when simulating the perturbed state trajectories from \eqref{eq:ypm}.

We use the Euler-Maruyama method to integrate the stochastic differential equations and obtain sample trajectories.  Euler-Maruyama is an extension of the forward Euler method for deterministic ordinary differential equations to the It\^{o} calculus.  A sample trajectory for $\Sigma_{S}$ is given by
\begin{equation}
X_{t+1} = X_t + f(X_t, u_t) \Delta t + \sigma(X_t, u_t) Z_t \sqrt{\Delta t}
\end{equation}
where each $Z_t \in \real^q$ is independently distributed as $\mathcal{N}(0,I)$ and $X_0$ is sampled from the initial condition distribution.  The sample $Y$ are then given by
\begin{equation}
Y_t = h(X_t).
\end{equation}
The sample trajectories of $y(t)$ can then be plugged into the definition of the empirical observability Gramian.
\begin{equation}
W_o^\eps(t_1, x_0, u) = \frac{1}{4\eps^2} \int_0^{t_1} \Phi^\eps(t,x_0,u)^T \Phi^\eps(t,x_0,u) dt,
\end{equation}
where
\begin{equation}
\Phi^\eps(t,x_0,u) = \begin{bmatrix}y^{+1}-y^{-1} & \cdots & y^{+n}-y^{-n}\end{bmatrix}
\end{equation}
and $y^{\pm i}(t)$ are independent sample trajectories of the system, $\Sigma_S$, with control input $u(t)$, initialized from $X_0 \pm \eps e_i$.  While $X_0$ itself is a random variable, we will assume that its distribution is described by the Dirac probability density function $\delta(X_0) = x_0$, that is, a single point.  This assumption can be relaxed, provided that the initial states of the sample trajectory are randomly chosen according to the desired initial condition distribution.

This modification results in a Gramian that is a random variable, and as a result, so are the unobservability index and estimation condition number.  We can numerically approximate the distribution of the observability indices by computing an ensemble of Gramians for a given initial condition.  As the estimation condition number and local unobservability index are both bounded below, it is important to note that the distributions of these variables will not be Gaussian.

Note that there are at least two reasonable ways to generate sample trajectories to compute a Gramian with stochastic dynamics.  The approach that we have taken here is to compute each perturbed trajectory, $y^{\pm i}$ once, requiring $2n$ simulations.  Another approach would be to compute new sample trajectories for each entry of the Gramian, requiring $4n^2$ simulations.  The choice of sampling techniques will influence the distribution of the Gramian, and its moments.  We will use the approach outlined in this section requiring fewer system evaluations in the rest of this paper.

\subsection{Expected value}
\label{sec:expectedvalue}
Now that the Gramian is a random variable, we are interested in what we can say about its moments.  In general, the moments are challenging to compute in closed-form, but we are able to derive some useful structure for the first moment in the general case.  In specific cases, namely linear stochastic systems with additive or multiplicative noise, we are able to compute the first moment of the Gramian more precisely.

Along the way, we shall need the following identity:

\begin{lemma}
\label{thm:expsqr}
For a random vector $X$ of length $n$,
\begin{equation}
\E\left[X^T X\right] = \E[X]^T \E[X] + \tr(\Cov[X])
\end{equation}
\end{lemma}
\begin{IEEEproof}
\begin{equation}
\begin{split}
\E\left[X^T X\right] & =  \E\left[\sum_{i = 0}^n X_i X_i\right]\\
& = \sum_{i = 0}^n \E[X_i X_i]\\
& = \sum_{i = 0}^n \E[X_i]^2 + \Var[X_i]\\
& = \E[X]^T \E[X] + \tr(\Cov[X])
\end{split}
\end{equation}
\end{IEEEproof}

We will also need a matrix version of the same lemma.  The $\diag(\cdot)_i$ operator that we use below is defined as mapping an $n$-dimensional vector, $v$ specified component-wise as $v_i$ to a diagonal $n \times n$ matrix whose diagonal elements are given by the components of $v$ and whose off-diagonal elements are all $0$.  In other words
\begin{equation}
\diag(v_i)_i = 
\begin{bmatrix}
v_1 & 0 & \hdots & 0 \\
0 & v_2 & & \vdots \\
\vdots & & \ddots & \\
0 & \hdots &  & v_n

\end{bmatrix}.
\end{equation}
\begin{lemma}
\label{thm:expsqrmat}
For a random matrix $X$, of length $n$, with independent columns
\begin{equation}
X = \begin{bmatrix}X_1 & X_2 & \dotsc & X_n\end{bmatrix},
\end{equation}
we have
\begin{equation}
\E\left[X^T X\right] = \E[X]^T \E[X] + \diag(\tr(\Cov[X_i]))_i
\end{equation}
\end{lemma}
\begin{IEEEproof}
\begin{equation}
\begin{split}
\E\left[X^T X\right] & =  \E
\begin{bmatrix}
X_1^T X_1 & X_1^T X_2 & \hdots & X_1^T X_n^T\\
X_2^T X_1 & X_2^T X_2 & & \vdots\\
\vdots & & \ddots &\\
X_n^T X_1 & \hdots & & X_n^T X_n
\end{bmatrix}\\
& = 
\begin{bmatrix}
\E[X_1^T X_1] & \E[X_1^T X_2] & \hdots &\E[ X_1^T X_n]\\
\E[X_2^T X_1] & \E[X_2^T X_2] & & \vdots\\
\vdots & & \ddots &\\
\E[X_n^T X_1] & \hdots & & \E[X_n^T X_n]
\end{bmatrix}.
\end{split}
\end{equation}
We can now apply Lemma~\ref{thm:expsqr} to each entry of the matrix and see that
\begin{equation}
\E\left[X^T X\right] = \E[X]^T \E[X] + \diag(\tr(\Cov[X_i]))_i,
\end{equation}
where we have used the independence of the columns to drop the covariance of the off-diagonal terms.
\end{IEEEproof}

We can now determine the expected value of the stochastic empirical observability Gramian as a function of the first two moments of the output trajectory distributions.

\begin{theorem}
\label{thm:meangram}
Let $\bar{W_o^\eps}$ be the matrix defined by 
\begin{equation}
\left(\bar{W_o^\eps}\right)_{ij} = \frac{1}{4\eps^2} \int_0^{t_1} \left(\E[y^{+i}] - \E[y^{-i}]\right)^T \left(\E[y^{+j}] - \E[y^{-j}]\right) dt
\end{equation}
and let $\hat{W}_o^\eps$ be the diagonal matrix defined by
\begin{equation}
(\hat{W}_o^\eps)_{ii} = \frac{1}{4\eps^2} \int_0^{t_1} \tr\left(\Cov[y^{+i}] + \Cov[y^{-i}]\right)dt.
\end{equation}
Then $\E[W_o^\eps(t_1, x_0, u)] = \bar{W}_o^\eps + \hat{W}_o^\eps$.
\end{theorem}
\begin{IEEEproof}
By definition,
\begin{equation}
(W_o^\eps)_{ij} = \frac{1}{4\eps^2} \int_0^{t_1} \left(y^{+i} - y^{-i}\right)^T \left(y^{+j} - y^{-j}\right) dt,
\end{equation}
where the $t_1$, $x_0$, and $u$ arguments have been dropped for brevity.  Taking the expectation on both sides, we get
\begin{equation}
\begin{split}
\E\left[(W_o^\eps)_{ij}\right] & = \E\left[\frac{1}{4\eps^2} \int_0^{t_1} \left(y^{+i} - y^{-i}\right)^T \left(y^{+j} - y^{-j}\right) dt\right]\\
& = \frac{1}{4\eps^2} \int_0^{t_1} \E\left[\left(y^{+i} - y^{-i}\right)^T \left(y^{+j} - y^{-j}\right)\right] dt.
\end{split}
\end{equation}
When $i \neq j$ the sample trajectories $y^{+i}$ and $y^{-i}$ are independent of $y^{+j}$ and $y^{-j}$, so for off-diagonal terms of the Gramian we get
\begin{equation}
\label{eq:offdiagterm}
\E[(W_o^\eps)_{ij}] = \frac{1}{4\eps^2} \int_0^{t_1} \hspace{-0.5em} \left(\E[y^{+i}] - \E[y^{-i}]\right)^T \hspace{-0.5em} \left(\E[y^{+j}] - \E[y^{-j}]\right) dt.
\end{equation}

Clearly, when $i = j$, independence does not hold.  By Lemma~\ref{thm:expsqr}, the diagonal terms of the Gramian become
\begin{multline}
\label{eq:diagterm}
\hspace{-1em} \E[(W_o^\eps)_{ii}] = \frac{1}{4\eps^2} \int_0^{t_1} \hspace{-0.5em} \left(\E[y^{+i}] - \E[y^{-i}]\right)^T \hspace{-0.5em} \left(\E[y^{+i}] - \E[y^{-i}]\right) dt\\ + \frac{1}{4\eps^2} \int_0^{t_1} \tr\left(\Cov[y^{+i} + y^{-i}]\right)dt.
\end{multline}
We can break these diagonal terms into two parts: one that depends on the variance of the samples and one that does not.  Note that the first term of \eqref{eq:diagterm} matches \eqref{eq:offdiagterm}, which is $(\bar{W}_o^\eps)_{ij}$.  The second term of \eqref{eq:offdiagterm} we can break down slightly further, by noting that $y^{+i}$ and $y^{-i}$ are independent, so that
\begin{equation}
\Cov[y^{+i} - y^{-i}] =\Cov[y^{+i}] - \Cov[y^{-i}].
\end{equation}
Therefore, the second term of \eqref{eq:diagterm} is just $(\hat{W}_o^\eps)_{ii}$.  We have now shown
\begin{equation}
\E[W_o^\eps(t_1, x_0, u)] = \bar{W}_o^\eps + \hat{W}_o^\eps.
\end{equation}
\end{IEEEproof}

For arbitrary nonlinear stochastic systems we cannot generally go any further than this theorem in closed-form, because for nonlinear measurement functions, $h(X)$, we cannot move the expectation inside the function, i.e. $\E[Y(t, x_0, u)] \neq h(\E[X(t, x_0, u])$.  However, when the output is linear ($Y = C X$), we can do so, and the $\bar{W}_o^\eps$ term becomes the Gramian of the expected trajectory.  For similar reasons, analytically computing the higher moments of the Gramian is not generally possible -- we cannot move the moment under the integral sign for arbitrary nonlinear systems.

Note that the second term of the expected Gramian contains all of the variance between the sample trajectories.  Each term of the expected Gramian is positive-semidefinite, and $\hat{W}_o^\eps$ is strictly positive definite whenever the system has no states that are decoupled from states with non-zero process noise input.  As a result, the $\hat{W}_o^\eps$ term, which captures much of the process noise influence, can only reduce the local unobservability index of the expected Gramian, though the estimation condition number can be increased or decreased by $\hat{W}_o^\eps$.

In practice, computing the expected Gramian from ensembles produced by Monte Carlo simulation may be more efficient than attempting to find the mean and covariance of the perturbed output trajectories themselves.  Such an approach would also allow the calculation of higher moments of the Gramian simultaneously. However, Theorem~\ref{thm:meangram} provides useful insight into the way in which process noise can influence and improve nonlinear observability.

In general, we cannot proceed further than this result in closed-form, because $\E[h(X)] \neq h(\E[X])$.  However, for linear stochastic systems with additive or multiplicative process noise, we can simplify this result further.

\subsubsection{Linear additive noise}

We can compute the expected value of Gramian for the non-scalar Ornstein-Uhlenbeck, though the result contains an integral term that cannot be completely evaluated analytically.  The non-scalar Ornstein-Uhlenbeck process is given by the SDE
\begin{equation}
\label{eq:ornsteinN}
\Sigma_{OU}:\quad
\begin{aligned}
dX & = A X dt + \Omega dW\\
Y & = C X.
\end{aligned}
\end{equation}
We can find the expected value of the Gramian for this process, by computing the first two moments of the measurement, $Y(t)$, using the linearity of the system.  The stochastic observability of systems of this type has been studied extensively \cite{Aoki1968,Chen1980,Mariton1986,Baram1987,West1994,Gustafsson2001,Hwang2003,Liu2009a,Liu2011,Shen2011,Subasi2014,Zhang2015b}, though in all but \cite{Mariton1986}, the analysis was in discrete-time.  In \cite{Hwang2003} the system being studied was a hybrid system with additional quantized states that altered the continuous state dynamics, and in \cite{West1994} the dynamics jumped between different linear systems at intervals given by a Markov chain.

\begin{corollary}
For the system $\Sigma_{OU}$,
\begin{equation}
\label{eq:meanOUN}
\begin{split}
\E[W_o^\eps] = & W_O + \frac{1}{2 \eps^2} \tr(W_O \Cov[X(0)])\\
& \quad + \frac{1}{2 \eps^2} \int_0^{t_1} \int_0^t \tr(C e^{A (t - \tau)} \Omega \Omega^T e^{A^T (t - \tau)} C^T) d\tau dt
\end{split}
\end{equation}
where $W_O$ is the deterministic linear observability Gramian for the system
\begin{equation}
\begin{aligned}
\dot x & = A x\\
y & = C x.
\end{aligned}
\end{equation}
\end{corollary}
\begin{IEEEproof}
For the first moment we have
\begin{equation}
\frac{d}{dt} \E[X] = A \E[X]
\end{equation}
which has the solution
\begin{equation}
\label{eq:ex}
\E[X(t)] = e^{A t} \E[X(0)].
\end{equation}
Therefore, by linearity of the expectation,
\begin{equation}
\E[Y^{\pm i}(t)] = C e^{A t} (\E[X(0)] \pm \eps e_i).
\end{equation}

To compute the variance of $Y(t)$, we first need $\E[X(t) X^T(t)]$, which has dynamics
\begin{equation}
\frac{d}{dt} \E[X X^T] = A \E[X X^T] + \E[X X^T] A^T + \Omega \Omega^T
\end{equation}
which is a Riccati equation.  The Riccati equation has a well-known solution (see \cite{Dorato2000} \S2.4)
\begin{equation}
\E[X X^T] = U_2 U_1^{-1}
\end{equation}
where $U_1, U_2 \in \real^{n \times n}$ are governed by
\begin{equation}
\frac{d}{dt} \begin{bmatrix}U_1 \\ U_2\end{bmatrix}
=
\begin{bmatrix}
-A^T & 0\\
\Omega \Omega^T & A
\end{bmatrix}
\begin{bmatrix}
U_1\\
U_2
\end{bmatrix}
\end{equation}
with initial conditions
\begin{equation}
\E[X(0) X^T(0)] = U_{2,0} U_{1,0}^{-1}.
\end{equation}
Note that these initial conditions give only $n^2$ equations for $2n^2$ unknowns, however, $U_{1,0}$ will cancel out of our final quantity and can be chosen arbitrarily, provided it is chosen to be invertible.

We solve straightforwardly for $U_1$
\begin{equation}
U_1 = e^{-A^T t} U_{1,0}
\end{equation}
which we can plug into the dynamics of $U_2$ to get
\begin{equation}
\begin{split}
\dot{U}_2 & = A U_2 + \Omega \Omega^T U_1\\
& = A U_2 + \Omega \Omega^T e^{-A^T t} U_{1,0}.
\end{split}
\end{equation}
Solving the linear system we get
\begin{equation}
\begin{split}
U_2 & = e^{A t} U_{2,0} + \int_0^t e^{A (t - \tau)} \Omega \Omega^T e^{-A^T \tau} U_{1,0} d\tau\\
& = e^{A t} U_{2,0} + e^{A t} \int_0^t e^{-A \tau} \Omega \Omega^T e^{-A^T \tau} d\tau  U_{1,0}.
\end{split}
\end{equation}
Therefore, we have
\begin{equation}
\label{eq:exx}
\begin{split}
\E[X X^T] & = e^{A t} U_{2,0} U_{1,0} e^{A^T t} + e^{A t} \int_0^t e^{-A \tau} \Omega \Omega^T e^{-A^T \tau} d\tau e^{A^T t}\\
& = e^{A t} (\E[X(0) X(0)^T] + \int_0^t e^{-A \tau} \Omega \Omega^T e^{-A^T \tau} d\tau) e^{A^T t}
\end{split}
\end{equation}

Because $Y$ is simply a linear function of $X$, we can write
\begin{equation}
\begin{split}
\Cov[Y] & = C \Cov[X] C^T\\
& = C (\E[X X^T] - \E[X] \E[X]^T) C^T.
\end{split}
\end{equation}
Substituting in \eqref{eq:ex} and \eqref{eq:exx} we find that the covariance of a sample measurement at time $t$ is given by
\begin{equation}
\begin{split}
\Cov[Y(t)] & = C e^{A t} (\E[X(0) X(0)^T] - \E[X(0)] \E[X(0)]^T\\
& \quad + \int_0^t e^{-A \tau} \Omega \Omega^T e^{-A^T \tau} d\tau) e^{A^T t} C^T\\
& = C e^{A t} \left(\Cov[X(0)]^T \right. \\
& \quad \left. + \int_0^t e^{-A \tau} \Omega \Omega^T e^{-A^T \tau} d\tau\right) e^{A^T t} C^T.
\end{split}
\end{equation}
Note that
\begin{equation}
\begin{split}
\Cov[X(0) \pm \eps e_i] & = \E[(X(0) \pm \eps e_i) (X(0) \pm \eps e_i)^T]\\
& \quad - \E[(X(0) \pm \eps e_i)] \E[(X(0) \pm \eps e_i)]^T\\
& = \E[X(0) X(0)^T] - \E[X(0)] \E[X(0)]^T\\
& = \Cov[X(0)].
\end{split}
\end{equation}
In other words, the covariance of the initial state distribution is not affected by perturbation.  Because the covariance of the measurement, $Y(t)$, depends linearly on the covariance of the state initial condition, we find that
\begin{equation}
\Cov[Y^{\pm i}(t)] = \Cov[Y(t)],
\end{equation}
i.e. the covariance of the measurement is also not affected by perturbations in the initial condition.

Now we can approach $\bar{W}_o^\eps$ and $\hat{W}_o^\eps$.  First, looking at the first term,
\begin{equation}
\begin{split}
(\bar{W}_o^\eps)_{ij} & = \frac{1}{4 \eps^2} \int_0^{t_1} (2 \eps C e^{A t} e_i)^T (2 \eps C e^{A t} e_j) dt\\
& = \int_0^{t_1} e_i^T e^{A^T t} C^T C e^{A t} e_j dt
\end{split}
\end{equation}
so that
\begin{equation}
\bar{W}_o^\eps = \int_0^{t_1} e^{A^T t} C^T C e^{A t} dt
\end{equation}
Note that this is the ordinary linear observability Gramian, which we called $W_O$.

The second term
\begin{equation}
\begin{split}
(\hat{W}_o^\eps)_{ii} & = \frac{1}{2 \eps^2} \int_0^{t_1} \tr(C e^{A t} (\Cov[X(0)] + \int_0^t e^{-A \tau} \Omega \Omega^T e^{-A^T \tau} d\tau) e^{A^T t} C^T) dt\\
& = \frac{1}{2 \eps^2} \int_0^{t_1} \tr( e^{A^T t} C^T \! C e^{A t} (\Cov[X(0)] + \int_0^t e^{-A \tau} \Omega \Omega^T e^{-A^T \tau} d\tau)) dt\\
& = \frac{1}{2 \eps^2} \tr(W_O \Cov[X(0)])\\
& \quad + \frac{1}{2 \eps^2} \int_0^{t_1} \!\!\!\! \int_0^t \!\! \tr(e^{A^T t} C^T \! C e^{A t} e^{-A \tau} \Omega \Omega^T e^{-A^T \tau}) d\tau dt\\
& = \frac{1}{2 \eps^2} \tr(W_O \Cov[X(0)])\\
& \quad + \frac{1}{2 \eps^2} \int_0^{t_1} \!\!\!\! \int_0^t \!\! \tr(C e^{A (t - \tau)} \Omega \Omega^T e^{A^T (t - \tau)} C^T) d\tau dt.
\end{split}
\end{equation}
The corollary then follows from Theorem~\ref{thm:meangram}.
\end{IEEEproof}

Note that \eqref{eq:meanOUN} can be re-written as 
\begin{equation}
\begin{split}
\E[W_o^\eps(t_1, x_0, u)] = & W_O(t_1) + \frac{1}{2 \eps^2} I \tr(W_O(t_1) \Cov[X(0)])\\
& + \frac{1}{2 \eps^2} I \int_0^{t_1} \tr(C W_C(t) C^T) d\tau dt,
\end{split}
\end{equation}
where $W_C(t)$ is the controllability Gramian of the linear system
\begin{equation}
\Sigma_\Omega:\quad
\begin{aligned}
\dot x & = A x + \Omega u\\
y & = C x,
\end{aligned}
\end{equation}
or as 
\begin{equation}
\begin{split}
\E[W_o^\eps(t_1, x_0, u)] = & W_O(t_1) + \frac{1}{2 \eps^2} I \tr(W_O(t_1) \Cov[X(0)]) \\
& + \frac{1}{2 \eps^2} I \int_0^{t_1} \tr(W_O(t) \Omega \Omega^T) d\tau dt.
\end{split}
\end{equation}
We can interpret $W_C(t_1)$ as noise transfer from control to state, meaning that stochastic observability is influenced by the noise-to-output power.

We also note that more noise (larger $\Omega$) never decreases the positive-definiteness of the expected Gramian, though the effect of noise on the estimation condition number is not necessarily monotonic.  In particular, the expected Gramian for $\Sigma_{OU}$ can be positive-definite even when the deterministic linear component of the system is not observable.  Furthermore, the effect of noise on the Gramian increases as $\eps \to 0$.  Increasing the initial covariance of the stochastic state, $\Cov[X(0)]$, also increases the positive-definiteness of the expected Gramian.  Because $\hat{W}_o^\eps \succcurlyeq 0$, the expected value of the Gramian for $\Sigma_{OU}$ will always be strictly positive-definite when the deterministic component of the system is observable.  If $W_O$ is not full rank, then $W_O$ and $\hat{W}_o^\eps$ must have non-intersecting null-spaces (except at the origin) in order for $\E[W_o^\eps]$ to be full rank.

\subsubsection{Multiplicative additive noise}
We now move to another stochastic variant of the classic LTI dynamics
\begin{equation}
\label{eq:bsN}
\Sigma_{BS}:\quad
\begin{aligned}
dX & = A X dt + \sum_{j = 1}^m \Omega_j X dw_j\\
Y & = C X,
\end{aligned}
\end{equation}
where the noise now depends multiplicatively on the state and $dw_j$ are the independent components of the process noise $dW$.  The stochastic observability of this system has been studied in continuous and discrete-time and usually with the addition of Markovian jumps between a finite set of dynamics ($A_i$, $\Omega_i$, $C_i$) in \cite{Zhang2004,Dragan2004,Dragan2006,Zhang2008a,Li2009,Sun2010,Ting2012,Shen2013}.  We will restrict ourselves to the simpler LTI case for this dissertation.

Note that one difference in these dynamics from the additive noise case discussed previously is that, once the system reaches equilibrium, it will remain there.  Because
\begin{equation}
\frac{d}{dt} \E[X] = A \E[X]
\end{equation}
we see that the system is stable in expectation when $A$ is Hurwitz.  In such a case we expect that the system state will eventually go to zero.

The $\diag(\cdot)_i$ operator that we use below is defined as mapping an $n$-dimensional vector, $v$ specified component-wise as $v_i$ to a diagonal $n \times n$ matrix whose diagonal elements are given by the components of $v$ and whose off-diagonal elements are all $0$.  In other words
\begin{equation}
\diag(v_i)_i = 
\begin{bmatrix}
v_1 & 0 & \hdots & 0 \\
0 & v_2 & & \vdots \\
\vdots & & \ddots & \\
0 & \hdots &  & v_n

\end{bmatrix}.
\end{equation}
We also define the operator $\vecop(\cdot)$, which maps an $n \times n$ matrix to an $n^2 \times 1$ vector by stacking the columns of the matrix, and the operator $\vecop^{-1}$, the inverse operation.  Note that each operator is linear.

\begin{corollary}
\label{thm:bsexpect}
For the system $\Sigma_{BS}$,
\begin{equation}
\label{eq:meanBSN}
\E[W_o^\eps(t_1, x_0, u)] = W_O(t_1) + \frac{1}{2} \int_0^{t_1} \diag \left( \tr \left( \vecop^{-1} \left( \omega_i \right) \right) \right)_i dt,
\end{equation}
where $W_O$ is the deterministic linear observability Gramian for the system
\begin{equation}
\begin{aligned}
\dot x & = A x\\
y & = C x,
\end{aligned}
\end{equation}
$Q = A \oplus A + \sum_{j = 1}^m \Omega_j \otimes \Omega_j$, and
\begin{equation}
\begin{split}
\omega_i = (C \otimes C) & \left( \frac{1}{\eps^2} e^{Q t} \vecop(\E[X(0) X(0)^T]) + e^{Qt} \vecop(e_i e_i^T) \right. \\
& \quad - \frac{1}{\eps^2} e^{(A \oplus A)t} \vecop(\E[X(0)] \E[X(0)]^T)\\
& \quad \left. - e^{(A \oplus A)t} \vecop(e_i e_i^T) \right).
\end{split}
\end{equation}
\end{corollary}
\begin{IEEEproof}
As before, for the first moment we have
\begin{equation}
\frac{d}{dt} \E[X] = A \E[X],
\end{equation}
which has the solution
\begin{equation}
\label{eq:exbs}
\E[X(t)] = e^{A t} \E[X(0)].
\end{equation}
Therefore, by linearity of the expectation,
\begin{equation}
\E[Y^{\pm i}(t)] = C e^{A t} (\E[X(0)] \pm \eps e_i).
\end{equation}

To compute $\hat{W}_o^\eps$, we need the covariance of $Y(t)$, for which we first need $\E[X(t) X^T(t)]$, which has dynamics
\begin{equation}
\frac{d}{dt} \E[X X^T] = A \E[X X^T] + \E[X X^T] A^T + \sum_{j = 1}^m \Omega_j \E[X X^T] \Omega_j^T.
\end{equation}
We can simplify this equation by making use of the identity $\vecop(\sum_{j = 1}^m \Omega_j \E[X X^T] \Omega_j^T) = \sum_{j = 1}^m (\Omega_j \otimes \Omega_j) \vecop(\E[X X^T])$.  Applying the identity, we get
\begin{equation}
\begin{split}
\frac{d}{dt} \vecop(\E[X X^T]) & = \vecop(A \E[X X^T]) + \vecop(\E[X X^T] A^T) \\
& \qquad + \sum_{j = 1}^m \vecop(\Omega_j \E[X X^T] \Omega_j^T)\\
& \hspace{-4em} = \left(A \otimes I + I \otimes A + \sum_{j = 1}^m \Omega_j \otimes \Omega_j \right) \vecop(\E[X X^T])\\
& \hspace{-4em} = \left(A \oplus A + \sum_{j = 1}^m \Omega_j \otimes \Omega_j \right) \vecop(\E[X X^T])\\
& \hspace{-4em} = Q \vecop(\E[X X^T]).
\end{split}
\end{equation}
Therefore,
\begin{equation}
\label{eq:exxbs}
\vecop(\E[X X^T]) = e^{Q t} \vecop(\E[X(0) X(0)^T]).
\end{equation}

Using the $\vecop$ identity again, we can write
\begin{equation}
\begin{split}
\vecop(\Cov[Y]) & = \vecop(C \Cov[X] C^T)\\
& = (C \otimes C) (\vecop(\E[X X^T]) - \vecop(\E[X] \E[X]^T)).
\end{split}
\end{equation}
Substituting in \eqref{eq:exbs} and \eqref{eq:exxbs} we find that the covariance of a sample measurement at time $t$ is given by
\begin{equation}
\label{eq:bscov}
\begin{split}
\vecop(\Cov[Y(t)]) = & (C \otimes C) \left( e^{Q t} \vecop(\E[X(0) X(0)^T]) \right.\\
& \left. - e^{(A \oplus A) t}\vecop(\E[X(0)]\E[X(0)]^T) \right)
\end{split}
\end{equation}
The covariance of the perturbed measurements is given by
\begin{equation}
\begin{split}
\vecop(\Cov[Y^{\pm i}(t)]) = & (C \otimes C) \left( e^{Q t} \vecop(\E[X(0) X(0)^T] \right. \\
& \pm \eps e^{Q t} \vecop(e_i \E[X(0)]^T) \\
& \pm \eps e^{Q t} \vecop(\E[X(0)] e_i^T)  \\
& + \eps^2 e^{Q t} \vecop(e_i e_i^T)\\
& - e^{(A \oplus A) t} \vecop(\E[X(0)]\E[X(0)]^T))\\
& \mp \eps e^{(A \oplus A) t} \vecop(e_i\E[X(0)]^T)))\\
& \mp \eps e^{(A \oplus A) t} \vecop(\E[X(0)] e_i^T) \\
& \left. - \eps^2 e^{(A \oplus A) t} \vecop(e_i e_i^T) \right) 
\end{split}
\end{equation}

Now we can solve for $\bar{W}_o^\eps$ and $\hat{W}_o^\eps$.  As before, the first term is given by
\begin{equation}
\begin{split}
(\bar{W}_o^\eps)_{ij} & = \frac{1}{4 \eps^2} \int_0^{t_1} (2 \eps C e^{A t} e_i)^T (2 \eps C e^{A t} e_j) dt\\
& = \int_0^{t_1} e_i^T e^{A^T t} C^T C e^{A t} e_j dt,
\end{split}
\end{equation}
giving
\begin{equation}
\begin{split}
\bar{W}_o^\eps & = \int_0^{t_1} e^{A^T t} C^T C e^{A t} dt\\
& = W_O(t_1).
\end{split}
\end{equation}

The $\hat{W}_o^\eps$ term has entries given by
\begin{equation}
\begin{split}
(\hat{W}_o^\eps)_{ii} = \frac{1}{2} \int_0^{t_1} \!\!\!\! \tr (\! & \vecop^{-1} ( (C \otimes C) (\frac{1}{\eps^2} e^{Q t} \vecop(\E[X(0) X(0)^T]) + e^{Qt} \vecop(e_i e_i^T)\\
& - \frac{1}{\eps^2} e^{(A \oplus A)t} \vecop(\E[X(0)] \E[X(0)]^T) - e^{(A \oplus A)t} \vecop(e_i e_i^T) ) ) ) dt
\end{split}
\end{equation}
The corollary then follows from Theorem~\ref{thm:meangram}.
\end{IEEEproof}

Unlike the case of the additive noise Ornstein-Uhlenbeck model, the condition number of the $\hat{W}_o^\eps$ matrix is not unity for $\Sigma_{BS}$.  When the noise depends on the state, the direction of the initial condition perturbation matters in the covariance of the output.  However, as before, the effect on the Gramian of noise increases as $\eps \to 0$, but in the case of multiplicative noise the effect is removed when the Gramian is evaluated with a Dirac delta initial condition at the origin, for which $\E[X(0)X(0)^T] = \E[X(0)]\E[X(0)]^T = 0$.

\subsection{Stochastic observability}
We are now ready to present the main result of this paper: a rank condition for the expected value of the empirical observability Gramian for stochastic observability.  The definition of stochastic observability that we use here originates with Dragan and Morozan \cite{Dragan2004} for linear stochastic systems.  Adapting that definition to our notation, we say that the system $\Sigma_{BS}$ is \emph{stochastically observable} if there exists $\beta > 0$ and $t_1 > 0$ such that
\begin{equation}
\label{eq:stochuniobsv}
\E\left[ \int_0^{t_1} \Psi^T(t,0) C(t)^T C(t) \Psi(t,0) dt \right] \succcurlyeq \beta I
\end{equation}
where $\Psi(t,t_0)$ is the fundamental matrix solution of $\Sigma_{BS}$.

Note that this definition is a slight simplification of the original, which applies to LTV systems with multiplicative noise and Markovian switches in the system matrices.  In our LTI, non-switching system, we can expand the left-hand side of \eqref{eq:stochuniobsv} in more detail.

The fundamental matrix solution of a stochastic linear system is itself a random variable, defined such that
\begin{equation}
X(t) = \Psi(t,0) X(0),
\end{equation}
i.e. the random variable of the state at time $t$ is the product of the random $\Psi(t,0)$ and the random initial state, which we will assume to be independently distributed.  We can derive the following properties of the random fundamental matrix,
\begin{align}
\E[X(t)] & = \E[\Psi(t,0)]\E[X(0)]\\
\Psi(t,t) & = I \quad \text{w.p.} 1.
\end{align}
We can also see that the columns of $\Psi(t,0)$ must be independent, because the $i$-th column is simply the solution of the system with initial condition $e_i$ and the solutions of the system from independent initial conditions must be independent.

Therefore, expanding and applying Lemma~\ref{thm:expsqrmat}, we get
\begin{equation}
\label{eq:stochobsv}
\begin{split}
& \E\left[ \int_0^{t_1} \Psi^T(t,0) C(t)^T C(t) \Psi(t,0) ds \right] \\
& \qquad = \int_0^{t_1} \E\left[ \Psi^T(t,0) C(t)^T C(t) \Psi(t,0) \right] dt\\
& \qquad = \int_0^{t_1} \E\left[ C(t) \Psi(t,0) \right]^T \E\left[C(t) \Psi(t,0) \right] dt \\
& \qquad \qquad + \int_t^{t + t_1} \diag\left( \tr \left( \Cov\left[ C(t) \phi(t,0) e_i \right] \right)\right)_i dt\\
& \qquad = \int_0^{t_1} e^{A^T t} C(t)^T C(t) e^{A t}  dt \\
& \qquad \qquad + \int_0^{t_1} \diag\left( \tr \left( \Cov\left[ C(t) \Psi(t,0) e_i \right] \right)\right)_i dt\\
& \qquad = W_O(t_1) + \int_0^{t_1} \diag\left( \tr \left( \Cov\left[ C(t) \Psi(t,0) e_i \right] \right)\right)_i dt.
\end{split}
\end{equation}
Taking a closer look at the second term, we get
\begin{equation}
\begin{split}
\Cov[C(t) \Psi(t,0) e_i] = & C(t) \Cov[\Psi(t,0) e_i] C(t)^T\\
= & C(t) \Cov[X^{+i}(t)] C(t)^T
\end{split}
\end{equation}
or, substituting from \eqref{eq:bscov},
\begin{equation}
\begin{split}
\vecop(\Cov[C(t) \Psi(t,0) e_i]) = & (C(t) \otimes C(t)) (e^{Q t} \vecop(e_i e_i^T)\\
& - e^{(A \oplus A) t} \vecop(e_i e_i^T)).
\end{split}
\end{equation}

We are now ready to state the main result of this section.

\begin{theorem}
\label{thm:stochobsv}
The system $\Sigma_{BS}$ is stochastically observable if and only if 
\begin{equation}
\rank(\E[W_o^\eps(t_1, \delta(0), 0)]) = n
\end{equation}
for some $t_1 > 0$ and any $\eps > 0$.
\end{theorem}
\begin{IEEEproof}
To begin, we will look at $\E[W_o^\eps(t_1, \delta(0), 0)]$.  Applying Corollary~\ref{thm:bsexpect}, we get
\begin{equation}
\begin{split}
\E[W_o^\eps(t_1, \delta(0), 0)] & = W_O(t_1) + \frac{1}{2} \int_0^{t_1} \diag( \tr ( \vecop^{-1} ( (C \otimes C) (e^{Qt} \vecop(e_i e_i^T) \\
& \qquad - e^{(A \oplus A)t} \vecop(e_i e_i^T) ) ) ) ) dt,
\end{split}
\end{equation}
which we note closely matches $\E\left[ \int_0^{t_1} \Psi^T(t,0) C(t)^T C(t) \Psi(t,0) dt \right]$ except for the factor of $\frac{1}{2}$.  In fact
\begin{equation}
\E\left[ \int_0^{t_1} \Psi^T(t,0) C(t)^T C(t) \Psi(t,0) dt \right] = \E[W_o^\eps] + \frac{1}{2} \hat{W}_o^\eps,
\end{equation}
where the arguments to the empirical observability Gramian have been dropped for clarity.

Now, if $\rank(\E[W_o^\eps]) = n$ for some $t_1 > 0$, then $\E[W_o^\eps] \succ 0$ because $W_o^\eps$ is symmetric and positive semi-definite by construction.

Applying Weyl's inequality, we get
\begin{equation}
\ubar{\lambda} \left(\E\left[ \int_0^{t_1} \Psi^T(t,0) C(t)^T C(t) \Psi(t,0) dt \right]\right) \geq \ubar{\lambda}(\E[W_o^\eps]) + \frac{1}{2} \ubar{\lambda}(\hat{W}_o^\eps).
\end{equation}

Let $\beta = \ubar{\lambda}(\E[W_o^\eps]) + \frac{1}{2} \ubar{\lambda}(\hat{W}_o^\eps)$.  We know that $\ubar{\lambda}(\E[W_o^\eps]) > 0$, and $\ubar{\lambda}(\hat{W}_o^\eps) \geq 0$, so it follows that $\beta > 0$.  Then
\begin{equation}
\E\left[ \int_0^{t_1} \Psi^T(t,0) C(t)^T C(t) \Psi(t,0) dt \right] \succcurlyeq \beta I,
\end{equation}
and the system is stochastically observable.

Now we consider the reverse case, and assume that the system is stochastically observable.  Assume that $\E[W_o^\eps]$ is not strictly positive definite.  It follows that neither $W_O$ or $\hat{W}^\eps_o$ can be strictly positive definite either, and that there exists at least one vector, $\eta$, that lies in the null spaces of both $W_O$ and $\hat{W}^\eps_o$.  However, in that case,
\begin{equation}
\begin{split}
\eta^T \! \E\left[ \int_0^{t_1} \!\!\! \phi^T(s,t) C^T C \phi(s,t) ds \right]\! \eta & = \eta^T \E[W_o^\eps] \eta + \frac{1}{2} \eta^T \hat{W}_o^\eps \eta\\
& = 0.
\end{split}
\end{equation}
However, that means that there can be no $\beta > 0$ that satisfies \eqref{eq:stochuniobsv}, which contradicts our assumption of stochastic observability.  Thus, by contradiction, $\E[W_o^\eps] \succ 0$, or $\rank(\E[W_o^\eps]) = n$.
\end{IEEEproof}

We note that the definition of stochastic observability from \cite{Dragan2004} does not readily extend to nonlinear systems, as it depends on the fundamental matrix, which has no analog in nonlinear systems.  The empirical observability Gramian extends naturally to nonlinear stochastic systems, however, meaning that the rank of the Gramian may potentially be used in a definition of stochastic nonlinear observability that is equivalent to existing definitions of both stochastic linear observability, deterministic linear observability, and, as shown in \cite{Powel2015}, partially equivalent to definitions of weak observability of deterministic nonlinear systems.

In Theorem~\ref{thm:stochobsv} the $\eps$ terms cancel out, just as they do in linear deterministic systems, but in the general nonlinear stochastic case, we cannot assume that this cancelation would take place.  In \cite{Powel2015}, the rank of the empirical Gramian in the limit as $\eps \to 0$ was used to demonstrate weak observability of nonlinear deterministic systems.  However, for nonlinear stochastic systems, the limit as $\eps \to 0$ does not exist.  As a result, a singular value condition, similar to Theorem~\ref{thm:empweakobsv}, proportional to $\eps^2 \tau$ would be a more appropriate way of defining stochastic nonlinear observability with the expectation of the empirical observability Gramian.

\section{Noise as modeling error}
\label{sec:err}

Noise enters our dynamics in a variety of ways.  Thermal fluctuations, aerodynamic turbulence, ambient electrical and radiological effects can all create noise.  These phenomena are also all examples of unmodeled dynamics, or modeling errors arising from simplifying approximations.  To understand how noise-as-modeling-error can influence the observability of a system, we create a simplified example system based on a linearization of an nonlinear system.

Consider the nonlinear system
\begin{equation}
\label{eq:nonlin}
\Sigma_{nl}:\quad
\begin{aligned}
\dot x & =
\begin{bmatrix}
-x_1 + \frac{1}{2} x_2^2\\
-x_2
\end{bmatrix}\\
y & = x_1,
\end{aligned}
\end{equation}
and its linearization at the equilibrium point $x = 0$
\begin{equation}
\label{eq:linearized}
\Sigma_{l}:\quad
\begin{aligned}
\dot x & =
\begin{bmatrix}
-1 & 0\\
0 & -1
\end{bmatrix}x\\
y & = \begin{bmatrix}1 & 0\end{bmatrix}x,
\end{aligned}
\end{equation}

From the Lie observability algebra of $\mathcal{O} = \{h, \Lie{f} h\}$ we find that the system $\Sigma_{nl}$ is observable when $x_2 \neq 0$, while clearly $\Sigma_l$ is nowhere observable.

Now consider the stochastic system
\begin{equation}
\label{eq:noiseaffine}
\Sigma_{sde}:\quad
\begin{aligned}
dx & = 
\begin{bmatrix}
-1 & 0\\
0 & -1
\end{bmatrix} x dt
+
\begin{bmatrix}
\frac{1}{2} x_2^2\\
0
\end{bmatrix} dW\\
y & = \begin{bmatrix}1 & 0\end{bmatrix}x,
\end{aligned}
\end{equation}
Note that the noise in this case is proportional to the modeling error in the linearization.  Numerically computing the empirical observability Gramian as above, we get an estimation condition number with a median value of 8.7 and an unobservability index with a median value of 20.5.  This value is compared to the deterministic estimation condition number of 22.5 and unobservability index of 40.2.

While this example may seem a bit contrived, it serves to illustrate how we can capture the influence on the observability of a nonlinear system of some kinds of unmodeled dynamic.  In particular, we have shown that an unobservable approximation to an observable system, can be seen to be observable when the noise is proportional to the approximation error.  Of course, in practice our noise model will never be exactly proportional to modeling error, because if we knew what the modeling error was, we could simply incorporate it into the model.  However, in many cases we can experimentally determine approximate noise characteristics to use in an approximate model, without knowing an exact model.  This approximate noise model could then be used in observability analysis of the approximate system.

\section{Simulation}
\label{sec:numeric}

To illustrate the use of the empirical observability Gramian with noise for observability of stochastic nonlinear systems, we have numerically computed ensembles of samples of the empirical observability Gramian for systems with process noise for two sample systems.  The first is a simple linear system, and the second a nonlinear unicycle model.  For each sample empirical observability Gramian, we computed the estimation condition number and the local unobservability index.

\subsection{Control/noise affine}

First we demonstrate a control and noise affine system system that is very nearly a linear system.  The system has dynamics
\begin{equation}
\Sigma_a:\quad
\begin{aligned}
dx & = \begin{bmatrix}-x_2 \\ x_1 u\end{bmatrix} dt + \begin{bmatrix}0\\ q x_1\end{bmatrix} dW\\
y & = x_2.
\end{aligned}
\end{equation}
First, we note that when $q = 0$ (no process noise), the system is observable if and only if $u(t) \neq 0$.  When $u(t) = 1$, we have a simple oscillator.

When the empirical observability Gramian is computed with no noise ($q = 0$) and control $u(t) = 0.1$, we find that the minimum eigenvalue of the Gramian is 0.497 and the condition number is 10.1.  As expected when $q= 0$ and $u(t) = 0$, the minimum eigenvalue is 0 and the condition number is undefined.  The system is linear for any particular constant control input, therefore these values are invariant to initial condition of the system.

\begin{figure}
\centering
\includegraphics[width=0.5\textwidth]{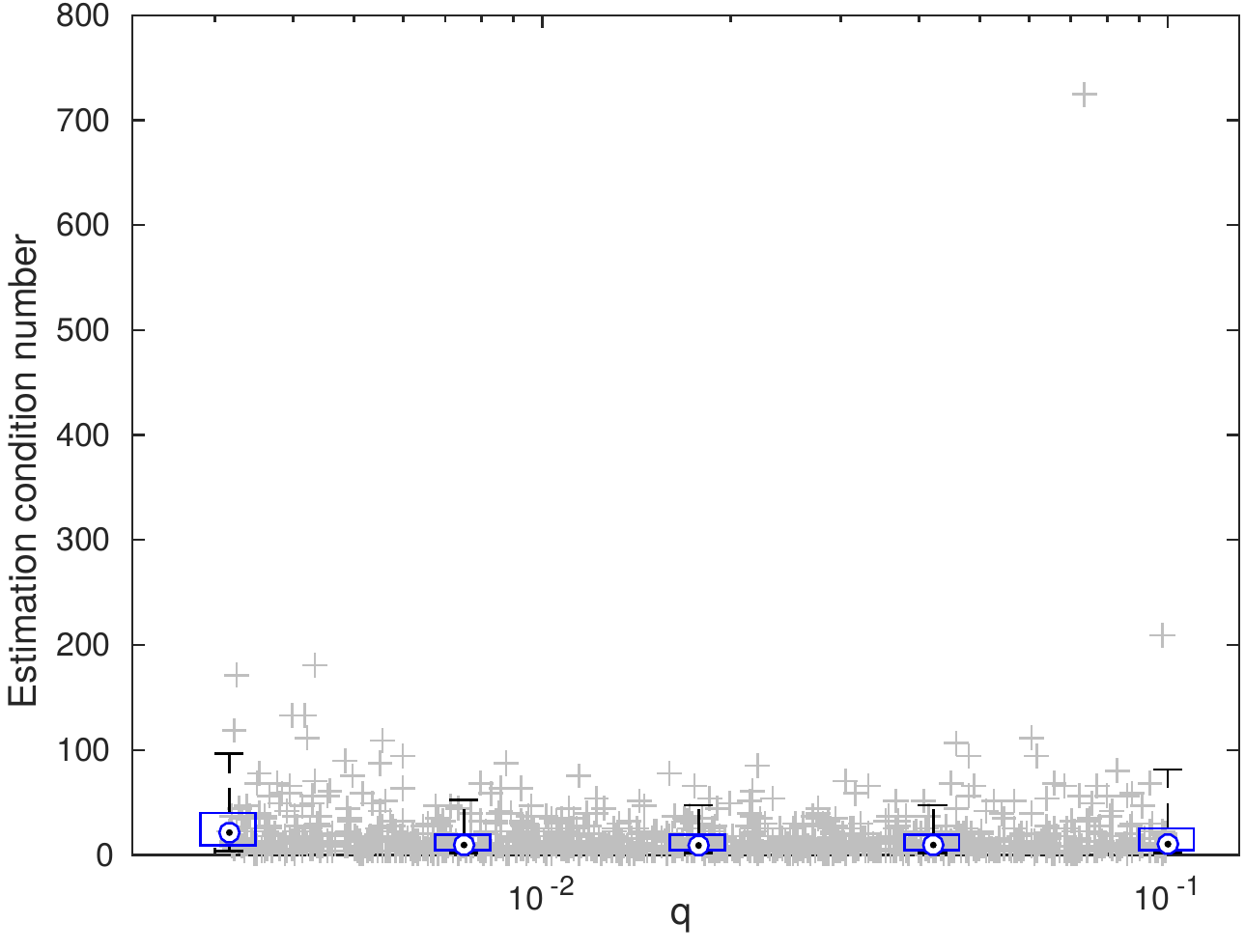}
\caption{Estimation condition number for the linear system shows a minimum near noise variance of $q=0.02$ for the noise affine system.  Too much and too little noise contribute to poorly conditioning.}
\label{fig:lincondnoise}
\end{figure}

\begin{figure}
\centering
\includegraphics[width=0.5\textwidth]{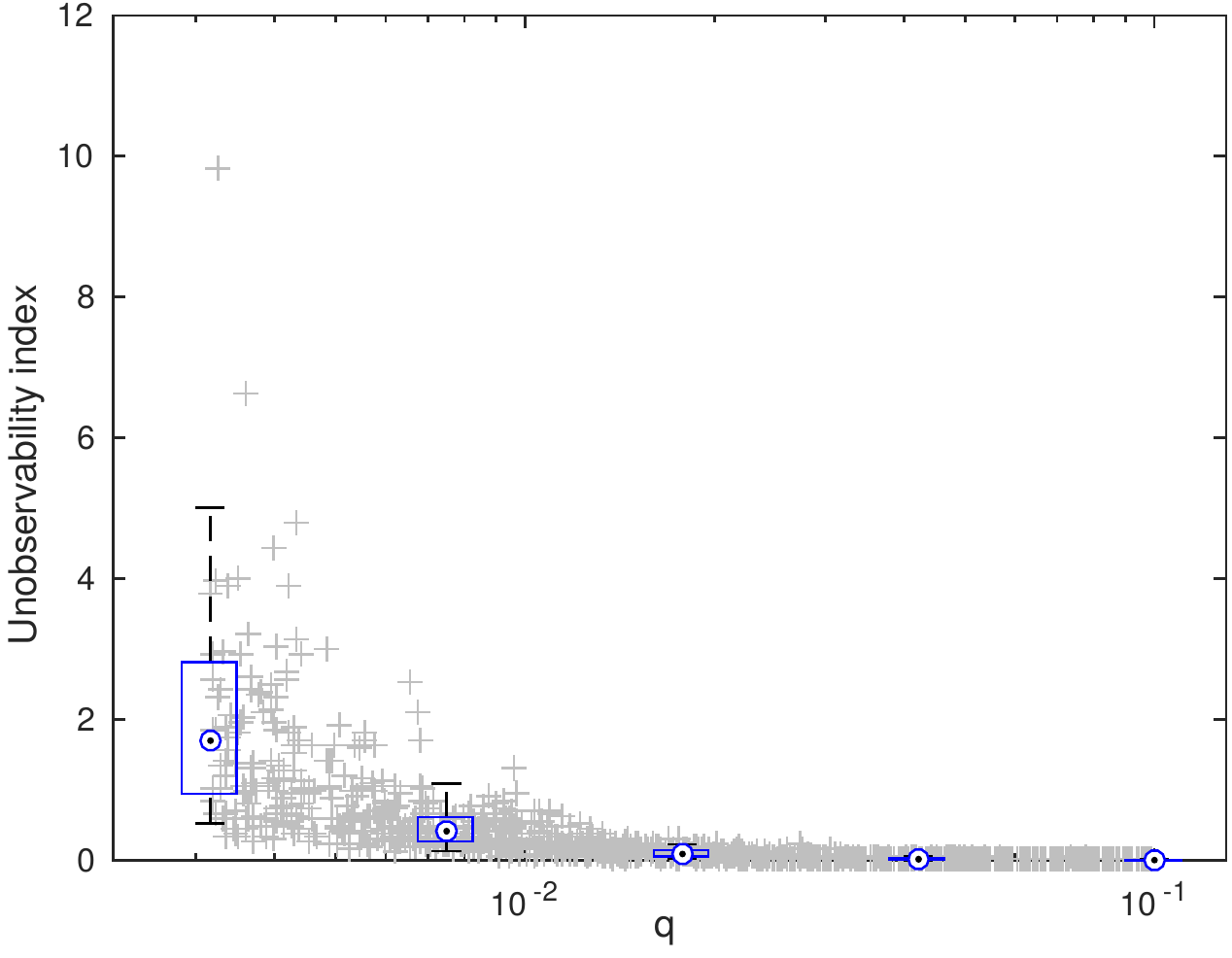}
\caption{Increasing noise appears to monotonically decrease the expected unobservability index in the noise affine system.}
\label{fig:lineignoise}
\end{figure}

If we consider the case $u(t) = 0$ and $q \neq 0$, we find that the condition number and minimum eigenvalue of the Gramian vary significantly depending on the sample trajectory.  Figure~\ref{fig:lincondnoise} and Figure~\ref{fig:lineignoise} show observability metrics for a range of $q$ values.  The gray points in the image are computed from Gramians computed with dynamics sampled uniformly from the $q$ domain, while the box plots are generated by an ensemble of 500 points at a fixed $q$.  The density of the gray points represent the distribution of the observability metrics over $q$ values and sample trajectories, while the box plots summarize the marginal distributions over the sample trajectories for a particular $q$.  The boxes cover the second and third quartiles, and the center dot shows the median.  The whiskers extend from the 5th to the 95th percentiles.


As the box plots show, the distribution of the metrics is strongly asymmetric, suggesting that mean and standard deviation are not the most pertinent descriptors of the values.  Instead, we will use the median as our primary summary statistic.

An important thing to note from the Figures~\ref{fig:lincondnoise} and \ref{fig:lineignoise} is that there appears to be a local minimum in the estimation condition number distributions as the noise variance is varied.  This minimum indicates that too little noise insufficiently actuates the system to produce observability (indeed, as $q \to 0$, the system becomes completely unobservable), while too much noise also impairs observability, perhaps by masking the actual dynamics of the system.

\begin{figure}
\centering
\includegraphics[width=0.5\textwidth]{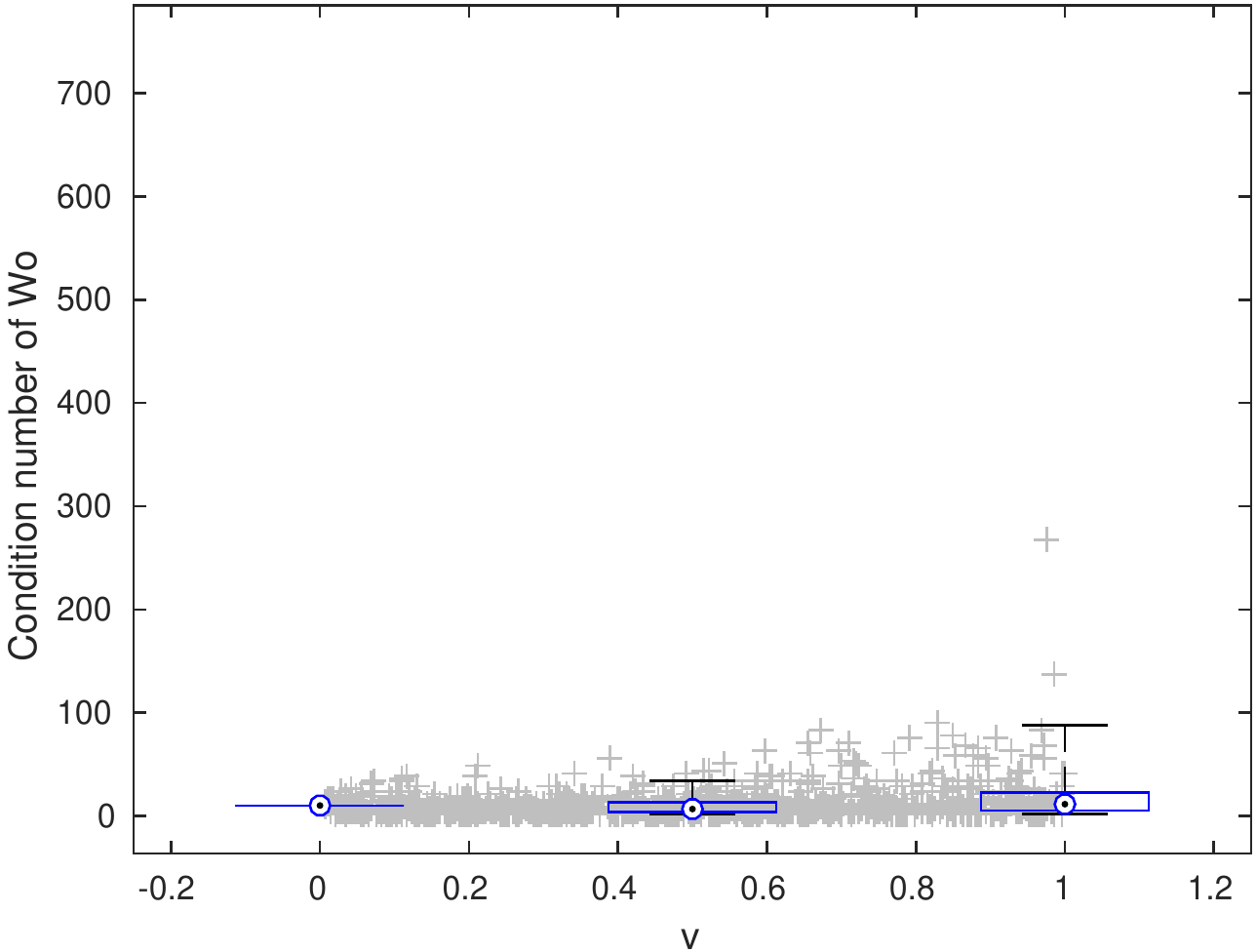}
\caption{Median estimation condition number improves slightly with mixed control and noise in the noise affine system, but pure control is better than pure noise.}
\label{fig:lincondcontrol}
\end{figure}

\begin{figure}
\centering
\includegraphics[width=0.5\textwidth]{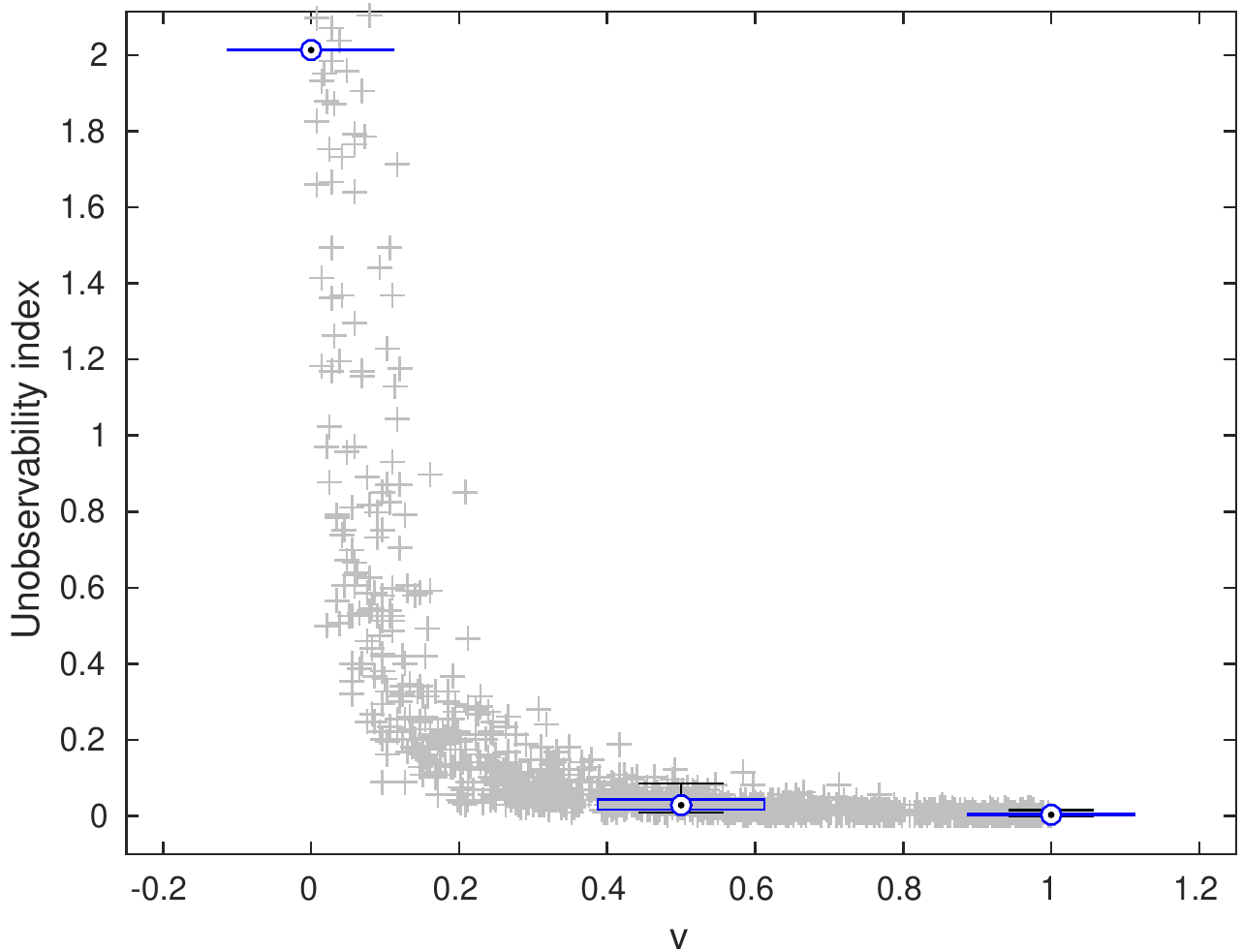}
\caption{Moving the balance from control to noise in the noise affine system provided an initial spike in the unobservability index, but provided a sharp decrease in the index overall.}
\label{fig:lineigcontrol}
\end{figure}

To compare the observability metrics resulting from control input and from noise, we computed the unobservability index and estimation condition number for the system
\begin{equation}
\Sigma_a:\quad
\begin{aligned}
dx & = \begin{bmatrix}-x_2 \\ x_1 (1 - v) u\end{bmatrix} dt + \begin{bmatrix}0\\ v q x_1\end{bmatrix} dW\\
y & = x_2.
\end{aligned}
\end{equation}
for $v \in [0, 1]$ for $u = 0.1$ and $q = 0.1$.  The parameter $v$ controls the trade-off between control and noise.  As Fig.~\ref{fig:lincondcontrol} and Fig.~\ref{fig:lineigcontrol} show, noise produced similar levels of observability to control for similar magnitudes, though with a much greater variation.


\subsection{Unicycle}

We can perform a similar set of computations for a nonlinear unicycle type vehicle with position measurement given by the dynamics
\begin{equation}
\Sigma_u:\quad
\begin{aligned}
dx & =
\begin{bmatrix}
x_4 \cos(x_3)\\
x_4 \sin(x_3)\\
u_1\\
u_2
\end{bmatrix} dt
+
\begin{bmatrix}
0\\
0\\
q\\
0
\end{bmatrix} dW
\\
y & =
\begin{bmatrix}
x_1\\
x_2
\end{bmatrix}.
\end{aligned}
\end{equation}
Note that the deterministic component of the system is fully observable only when the vehicle is moving, $x_4 \neq 0$, or accelerating, $u_2 \neq 0$.  As before, we can compute Gramians for the system at equilibrium, $x = 0$, and with no control input, $u(t) = 0$, across a range of noise values.  We expect in this case, that noise will cause the system to become observable and we intuitively justify this expectation by noting that noise in the speed state will result in the vehicle jittering along the heading direction of the vehicle, providing information in the output about that state that would not otherwise be available.

\begin{figure}
\centering
\includegraphics[width=0.5\textwidth]{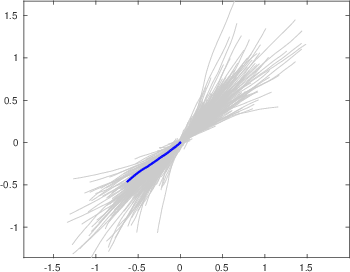}
\caption{1000 sample runs with one highlighted run illustrate that acceleration noise can render the initial unicycle heading of $45^\circ$ observable up to $180^\circ$.}
\label{fig:intuition}
\end{figure}

\begin{figure}
\centering
\includegraphics[width=0.5\textwidth]{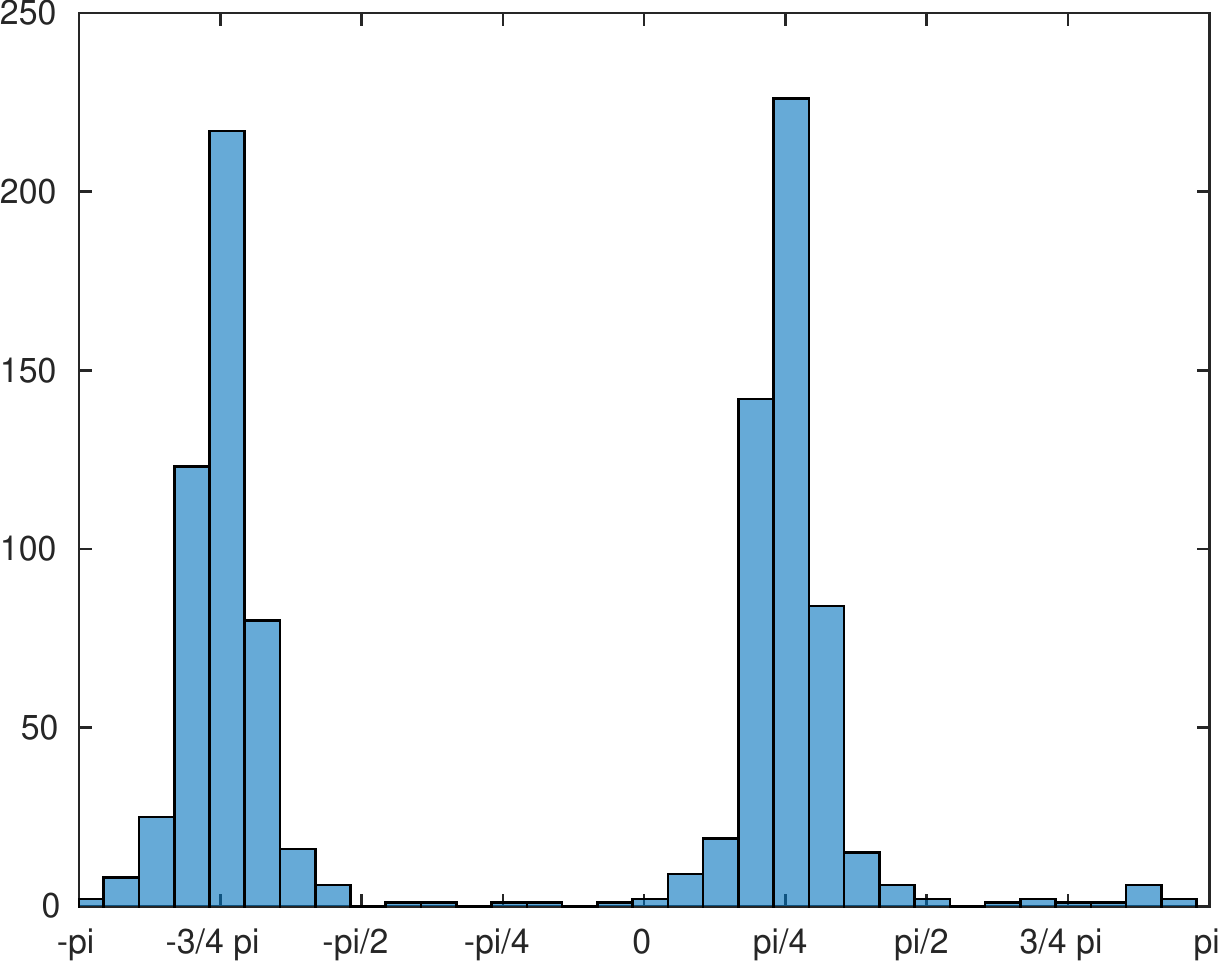}
\caption{The forward and reverse headings are clearly visible in the histogram of directions of the final position of the vehicle from the origin.}
\label{fig:intuitdist}
\end{figure}

Figure~\ref{fig:intuition} and Figure~\ref{fig:intuitdist} demonstrates this intuition by showing the measurement trajectories of a unicycle across 1000 sample runs, each starting at the origin with an initial heading of $45^\circ$.  The highlighted trajectory from Figure~\ref{fig:intuition} shows that for any particular sample, we can more-or-less judge the initial heading of the vehicle, which deterministic observability analysis would not have told us was possible.  Figure~\ref{fig:intuitdist} shows that in ensemble, the initial heading can be seen in a maximum likelihood sense, up to the forward/backward ambiguity.  Ensemble conclusions are probably not as useful in most situations, as they require multiple runs from the system, but they can be useful to illustrate our intuition in this scenario.

\begin{figure}
\centering
\includegraphics[width=0.5\textwidth]{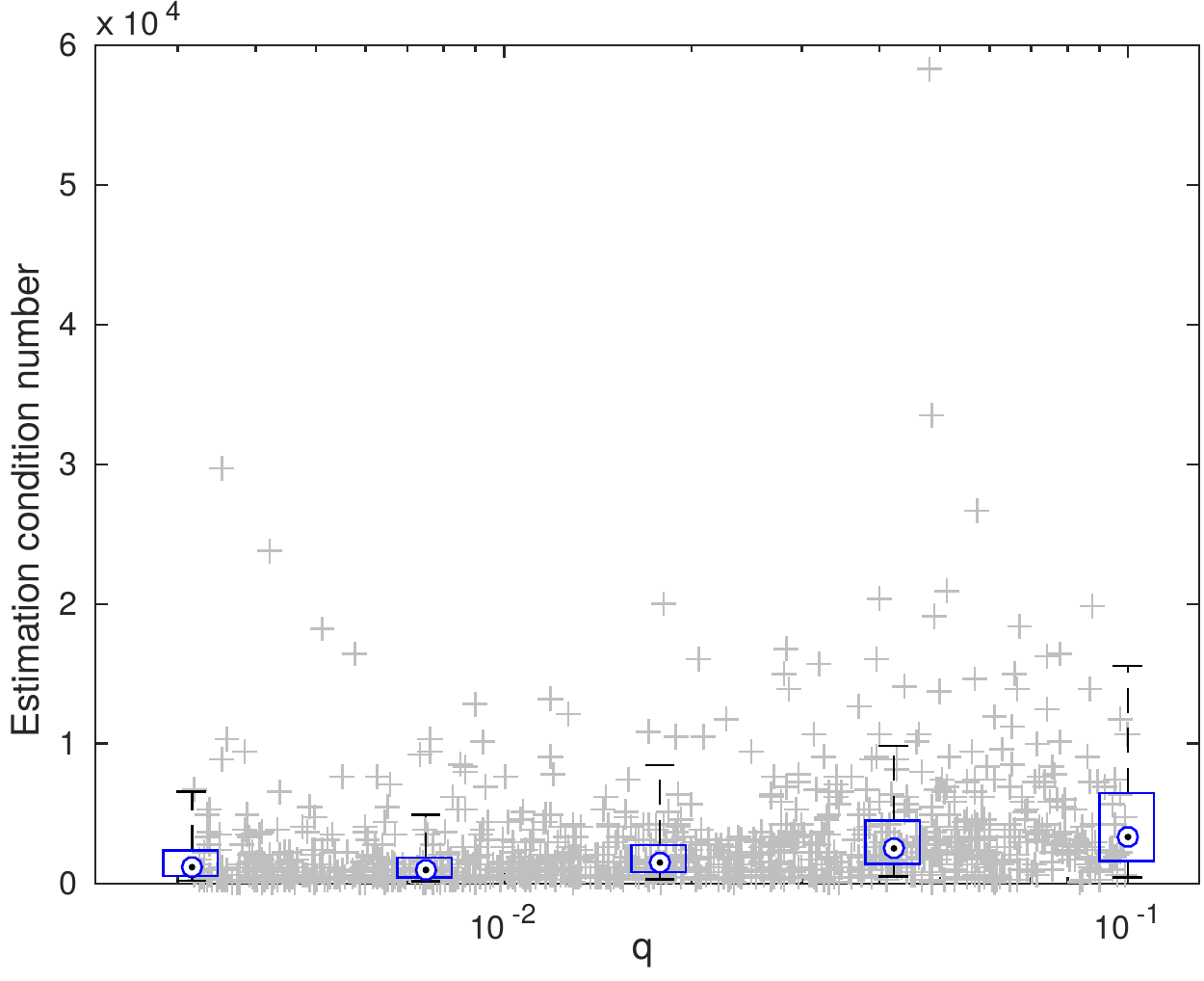}
\caption{Estimation condition number for the unicycle system has a minimum near $q = 0.008$.  As before, too much and too little noise contribute to poor conditioning.}
\label{fig:unicondnoise}
\end{figure}

\begin{figure}
\centering
\includegraphics[width=0.5\textwidth]{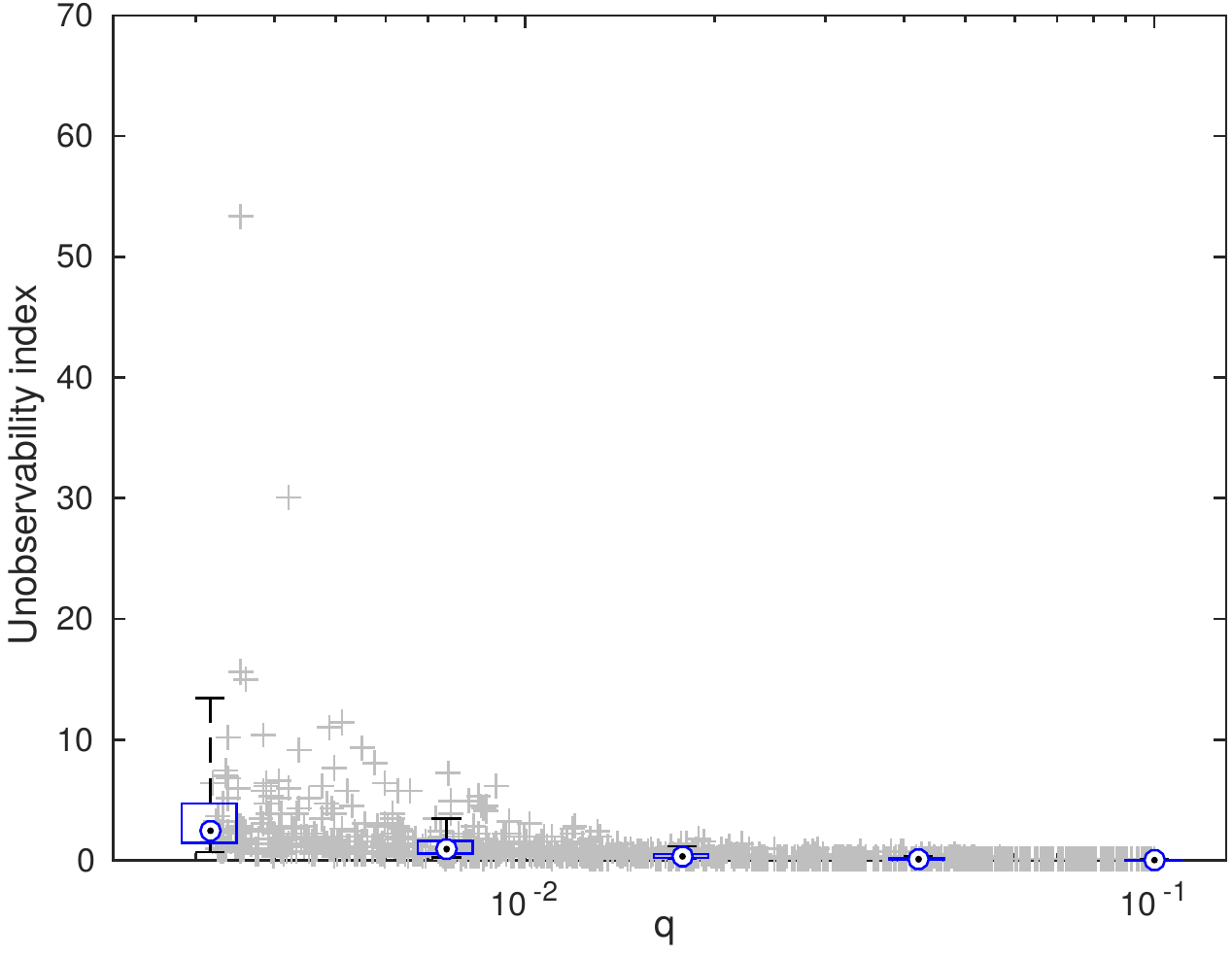}
\caption{Increasing noise appears to monotonically decrease the unobservability index for the unicycle system.}
\label{fig:unieignoise}
\end{figure}

\begin{figure}
\centering
\includegraphics[width=0.5\textwidth]{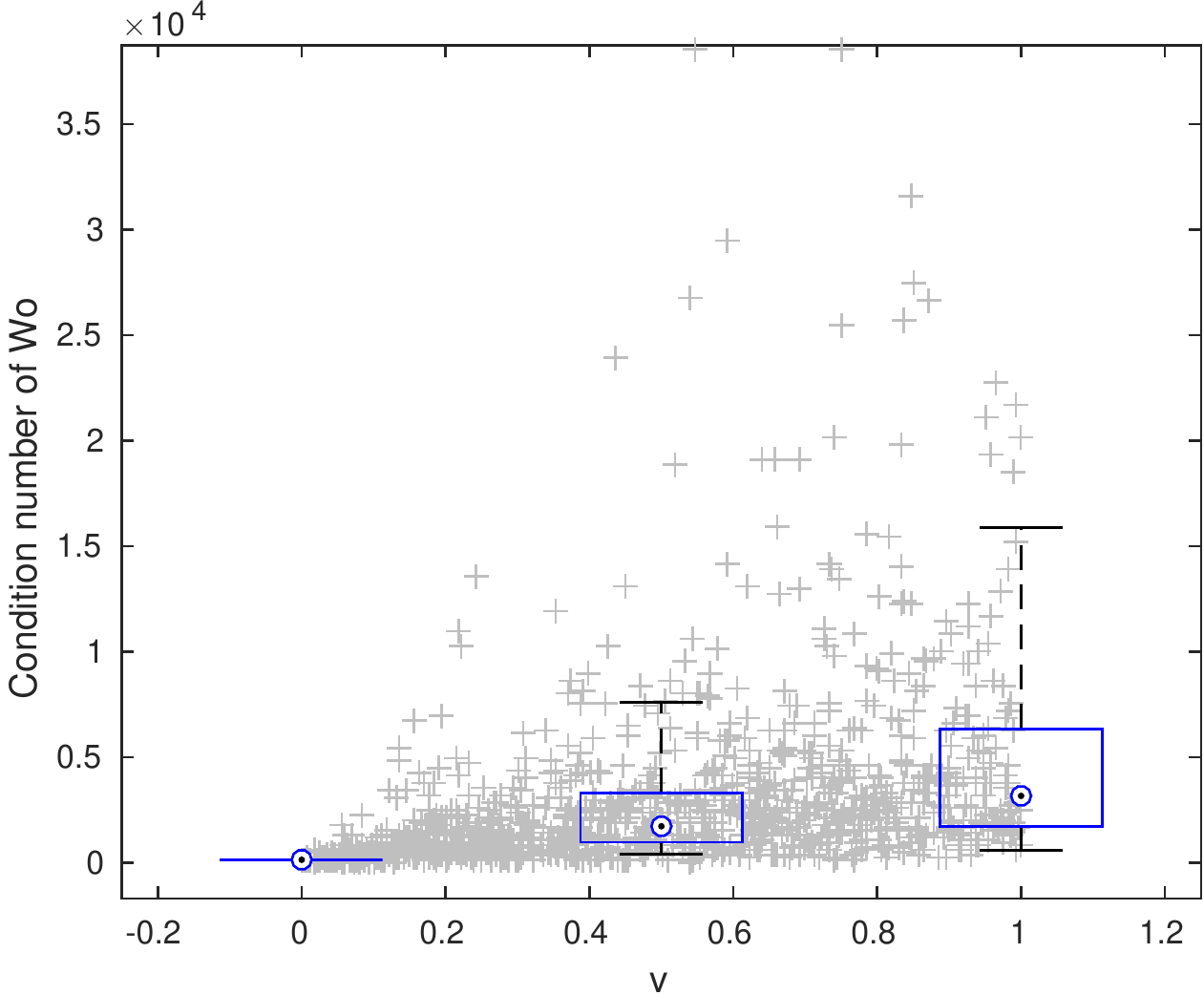}
\caption{Control provided better conditioning than mixed or pure noise for the unicycle.}
\label{fig:unicondcontrol}
\end{figure}

\begin{figure}
\centering
\includegraphics[width=0.5\textwidth]{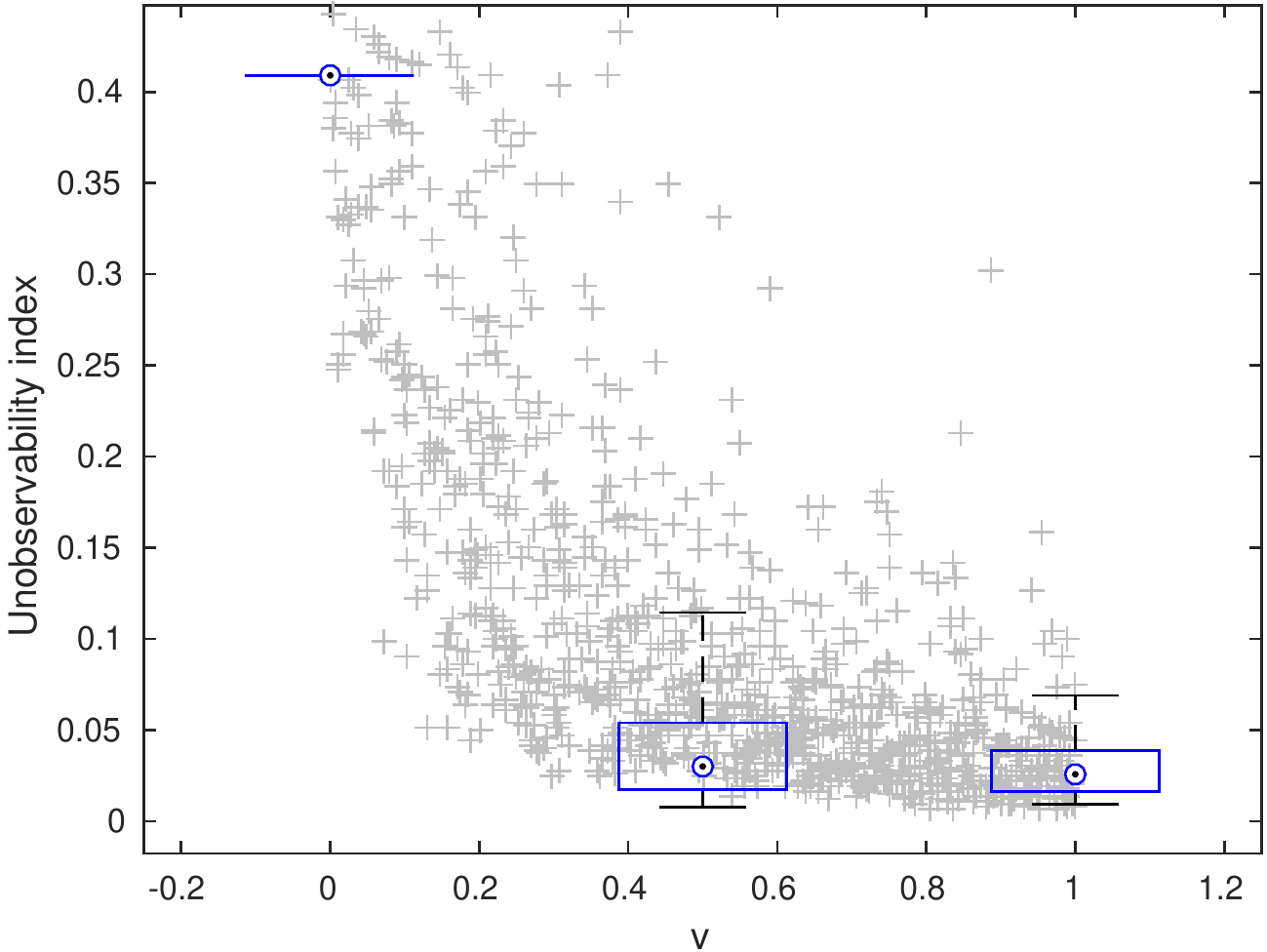}
\caption{Increasing noise appears to monotonically decrease the unobservability index in the unicycle system.}
\label{fig:unieigcontrol}
\end{figure}

Figures~\ref{fig:unicondnoise}-\ref{fig:unieigcontrol} show the observability metrics we computed for a range of $q$ and, as before, for a trade-off between control and noise.  As before, we find a local minimum in the estimation condition number.  Note that the condition number can range quite high (to very poorly conditioned estimation) values, but that the median condition number stays quite close to the condition number with pure control.



While the noise analysis here might seem superfluous, given that control was available to induce observability in the same parts of the dynamics as the noise, in general, this situation need not be the case.  We structured these systems to have similar noise and control inputs so that noise and control influences on observability could be compared, but in many systems, noise and control enter the dynamics in very different ways.  Noise, in general, has the potential to induce observability of states that cannot be excited by the available controls.

It may appear that in this example we have added noise only to the state calculated to provide the most benefit.  However, this scenario may be physically reasonable if we are considering, for example, electrical noise in a drive motor.  Furthermore, the intuition that leads us to this experiment still holds in general, even if we add noise to other states (depending on magnitude, of course).  There is no reason to believe that noise is always going to affect each state equally, and if there were some noise in the steering, our noisy measurements of position would now, instead of lying on a line, as before, lie in the space shown in Figure~\ref{fig:intuition}, the major axis of which will still provide information about the heading of the vehicle.   Even assuming that the steering noise exceeded the acceleration noise, we can always back out what we want as long as we don't get a circle and if we know the relative magnitudes of the noises.  We do not consider noise position states for this system, because they are related kinematically to the heading/velocity states.

\section{Conclusions}
\label{sec:conclusions}

We have demonstrated that process noise can impact the observability metrics of linear and nonlinear systems, and even cause deterministically unobservable systems to become observable.

While noise can induce observability in unobservable systems, this property does not immediately lead to straightforward improvements in estimation or estimators.  For example, while the noise in the nonlinear unicycle model caused the otherwise unobservable heading state to appear in the output of the system, a traditional estimator will still struggle to estimate the vehicle heading, because the sign of the noise at any given instant is unknown, making forward and reverse motion indistinguishable.  In that case, the noise allows us to estimate the heading $\pm \pi$, but not the exact heading.  Extending the observability results from this work into improved estimators is beyond the scope of this paper.

At this point we are neglecting noise in the measurement for simplicity.  While the inclusion of sensor noise into the numerical results is simple, further work needs to be done to develop the analytically results with measurement noise included.  Another avenue of future research would be to investigate connections between the empirical Gramian with noise and definitions for stochastic observability such as that of Liu and Bitmead \cite{Liu2011}.  Another avenue of investigation would be the connection between the empirical observability Gramian (both deterministic and stochastic) and Lie algebras as they connect to nonlinear observability.


\bibliographystyle{IEEEtran}
\bibliography{IEEEabrv,Powel-EmpiricalGramian}

\newpage

\begin{IEEEbiographynophoto}{Nathan Powel}
Nathan D. Powel received the B.S. (summa cum laude) in Mathematics and in Aeronautics and Astronautics from the University of Washington, Seattle, WA, USA in 2009, and received the M.S. and Ph.D. in Aeronatics and Astronautics in 2011 and 2016 respectively from the University of Washington.

His research interests include the observability of nonlinear systems and the interaction of process noise with nonlinear dynamics.  He currently works in the aerospace industry in the development of manned and reusable space systems.
\end{IEEEbiographynophoto}

\vspace{-6.5in}

\begin{IEEEbiographynophoto}{Kristi Morgansen}
Kristi A. Morgansen (SM'06) received the B.S. (summa cum laude) and the M.S. in Mechanical Engineering from Boston University, Boston, MA, USA respectively in 1993 and 1994, and the S.M. in Applied Mathematics and Ph.D. in Engineering Sciences respectively in 1996 and 1999 from Harvard University, Cambridge, MA, USA.

She is currently Professor in the William E. Boeing Department of Aeronautics \& Astronautics at the University of Washington in Seattle, WA, USA. From 2002 to 2007, Professor Morgansen held the chaired position of Clare Boothe Luce Assistant Professor of Engineering at the University of Washington. She received a National Science Foundation (NSF) CAREER Award in 2003 and the O. Hugo Schuck award in the Theory category in 2010.  She is an Associate Fellow of the AIAA.   Her research interests focus on nonlinear systems where sensing and actuation are integrated, stability in switched systems with delay, and incorporation of operational constraints such as communication delays in control of multi-vehicle systems.  Applications include both traditional autonomous vehicle systems as well as bio-inspired and biological underwater and flight systems, human decision making, and neural engineering.
\end{IEEEbiographynophoto}

\end{document}